\documentclass[pdflatex,sn-nature]{sn-jnl}

\usepackage{graphicx}
\usepackage{amsmath,amssymb,amsfonts}
\usepackage{xcolor}
\usepackage{caption}
\usepackage{hyperref}
\usepackage{braket}

\unnumbered

\begin{document}

\title{Towards dislocation-driven quantum interconnects}

\author[1]{\fnm{Cunzhi} \sur{Zhang}}
\equalcont{These authors contributed equally to this work.}

\author[2]{\fnm{Victor Wen-zhe} \sur{Yu}}
\equalcont{These authors contributed equally to this work.}

\author[1]{\fnm{Yu} \sur{Jin}}

\author[1]{\fnm{Jonah} \sur{Nagura}}

\author[3]{\fnm{Sevim Polat} \sur{Genlik}}

\author*[3]{\fnm{Maryam} \sur{Ghazisaeidi} \email{ghazisaeidi.1@osu.edu}}

\author*[1,2,4]{\fnm{Giulia} \sur{Galli} \email{gagalli@uchicago.edu}}

\affil[1]{\orgdiv{Pritzker School of Molecular Engineering}, \orgname{University of Chicago}, \orgaddress{Chicago, IL 60637, USA}}
\affil[2]{\orgdiv{Materials Science Division}, \orgname{Argonne National Laboratory}, \orgaddress{Lemont, IL 60439, USA}}
\affil[3]{\orgdiv{Department of Materials Science and Engineering}, \orgname{The Ohio State University}, \orgaddress{Columbus, OH 43210, USA}}
\affil[4]{\orgdiv{Department of Chemistry}, \orgname{University of Chicago}, \orgaddress{Chicago, IL 60637, USA}}

\abstract{
A central problem in the deployment of quantum technologies is the realization of robust architectures for quantum interconnects. We propose to engineer interconnects in semiconductors and insulators by patterning spin qubits at dislocations, thus forming quasi one-dimensional lines of entangled point defects. To gain insight into the feasibility and control of dislocation-driven interconnects, we investigate the optical cycle and coherence properties of nitrogen-vacancy (NV) centers in diamond, in proximity of dislocations, using a combination of advanced first-principles calculations. We show that one can engineer spin defects with properties similar to those of their bulk counterparts, including charge stability and a favorable optical cycle, and that NV centers close to dislocations have much improved coherence properties. Finally, we predict optically detected magnetic resonance spectra that may facilitate the experimental identification of specific defect configurations. Our results provide a theoretical foundation for the engineering of one-dimensional arrays of spin defects in the solid state.
}

\maketitle

\section{Introduction}

One of the key components of quantum technologies is the interconnect -- an architecture or physical process that enables the transfer of information between disparate and varied quantum nodes~\cite{kimble_quantum_2008,atature_material_2018,wehner_quantum_2018,awschalom_development_2021}. In the solid state, controlling and scaling defect-based qubits into coherent, interconnected arrays remain major obstacles, limiting the application of spin defects for quantum technologies~\cite{smith_colour_2019,castelletto_silicon_2020,wolfowicz_quantum_2021,chatterjee_semiconductor_2021}. Here, we address this challenge and propose to create quantum interconnects by using the quasi one-dimensional (1D) atomic arrangements present at dislocations to pattern and eventually entangle spin qubits in semiconductors and insulators.

Dislocations are topological, line-defects in crystals~\cite{anderson_theory_2017}. At their narrow core, atoms adopt complex arrangements that differ from those in the rest of the solid. The unique bonding environments of dislocation cores can give rise to emergent electronic and magnetic properties~\cite{nakamura_conducting_2003,sugiyama_ferromagnetic_2013,ran_one_2009,genlik_dislocations_2023}; in addition, dislocations can attract vacancies and substitutional impurities~\cite{bullough_kinetics_1970,ghassemizadeh_stability_2022,yani_assessing_2024}, due to their long-range stress/strain field, offering an opportunity to engineer unique structures of interacting line and point defects.

Recently, some preliminary theoretical studies explored the prospect of assembling arrays of spin defects~\cite{ghassemizadeh_stability_2022,yani_assessing_2024}. Close to the core of dislocations, the formation energies of nitrogen vacancy centers (NV) in diamond~\cite{ghassemizadeh_stability_2022} and divacancies in silicon carbide~\cite{yani_assessing_2024} were found to be much lower relative to the bulk (up to 3-4 eV lower). Interestingly, Ref.~\citenum{ghassemizadeh_stability_2022} identified an NV configuration at the 30$^{\circ}$ dislocation core that exhibits stability properties similar to the corresponding bulk NV. These results suggest that spin defects may be patterned to form interconnects of qubits, exploiting the thermodynamic driving force provided by dislocations.

However, the viability of spin defects as qubits close to dislocations, and specifically their optical cycle and coherence properties, has never been investigated, and the impact of line defects on their functionality remains unknown. In fact, it is only recently that advanced computational methods to study the full optical cycles of NV centers, including inter-system crossing (ISC) rates, became available~\cite{jin_first_2025} and that calculations beyond mean-field, density functional theory (DFT) could be applied to systems complex enough to describe, at the same time, dislocations and point defects~\cite{yu_gpu_2024}. To provide a theoretical foundation for the engineering of 1D arrays of spin defects in the solid state, we present a detailed investigation of the optical cycle of NV centers close to dislocations as a representative system, using a combination of recently developed first-principles methods and large-scale simulations. Our investigation integrates electronic structure and coherence time calculations, uncovering novel dynamical properties of spin defects close to dislocations, in addition to providing predictions on how to leverage the interaction of spin defects with line dislocations to enhance their properties.

Through a high-throughput search of stable centers near specific dislocations, we identify triplet configurations stable in the negative (i.e., $-1$) charge state. We then carry out an in-depth investigation of the optical cycle of representative NVs. We show that despite the variability of properties depending on the geometrical configuration, one can engineer spin defects near dislocations that not only retain the essential characteristics required for qubit functionality, but also exhibit longer coherence times compared to those in pristine diamond.

\section{Results}

\subsection{NV centers close to dislocation cores}

We focus on the 30$^{\circ}$ and 90$^{\circ}$ glide partial dislocations, which arise from the energetically favorable dissociation of the 60$^{\circ}$ glide perfect dislocations~\cite{blumenau_dislocations_2002,blumenau_dislocations_2003}. Specifically, we consider the double-period reconstruction of the core geometry~\cite{genlik_dislocations_2023}, which reduces the energy of the system by minimizing the number of dangling bonds (Fig.~S1). We carry out first-principles calculations with a combination of advanced techniques, ranging from DFT and time-dependent (TD) DFT with hybrid functionals and analytical forces~\cite{jin_excited_2023} to quantum defect embedding~\cite{sheng_green_2022,chen_advances_2025} and cluster-expansion for coherence times calculations~\cite{yang_quantum_2008,onizhuk_probing_2021}. We start by describing the stability of NV centers, and then we discuss their electronic properties and optical cycle.

\begin{figure}[!ht]
\centering
\includegraphics[width=0.8\textwidth]{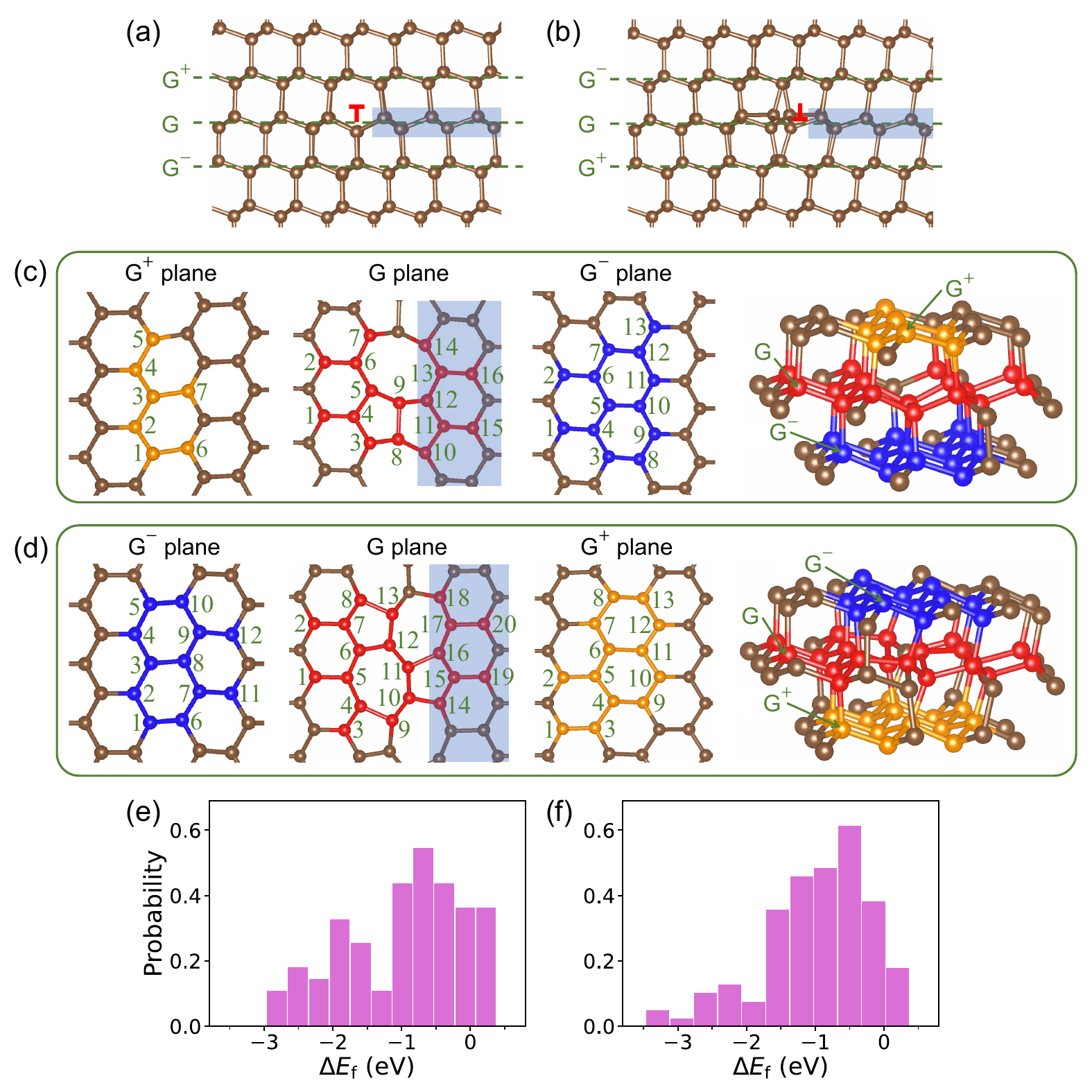}
\caption{
\textbf{Atomic sites chosen to create NV defects near dislocation cores in diamond.}
(a) Structure of the 30$^{\circ}$ dislocation in diamond. The \{111\} planes are indicated by dashed green lines near the dislocation core marked by the red $\top$ symbol. The shaded region indicates the stacking fault. The glide plane of the dislocation is denoted as the G plane; the adjacent \{111\} plane under compression (tension) is denoted as G$^-$ (G$^+$). (b) Same as (a), but for the 90$^{\circ}$ dislocation.
(c) Atomic sites considered to create NV defects near the 30$^{\circ}$ dislocation core. NV defects are created in the G, G$^-$, and G$^+$ \{111\} planes, with the N atom and the nearby carbon vacancy located at one of the orange, red, or blue atomic sites. We denote an atomic site based on the \{111\} plane it resides in and the number assigned in the figure. For instance, G$^+$1 stands for the atomic site 1 in the G$^+$ plane. (d) Same as (c), but for the 90$^{\circ}$ dislocation.
(e) Distribution of relative formation energies ($\Delta E_{\rm f}$, obtained using DFT at the PBE level of theory; see SI) of the 90 NV configurations near the 30$^{\circ}$ dislocation core compared to that in bulk diamond. (f) Same as (e), but for the 112 NV configurations near the 90$^{\circ}$ dislocation core.}
\label{fig:NV_location}
\end{figure}

Fig.~\ref{fig:NV_location} (a-d) shows the core structures of two dislocations and the atomic sites considered to create NV defects near the core region, and (e-f) the formation energy of all NV configurations analyzed here [see Supplementary Information (SI) for details of the enumeration of NV configurations]. In total, we screened 90 (112) NV arrangements near the 30$^{\circ}$ (90$^{\circ}$) dislocation core considering all atomic sites colored in orange, red, and blue in Fig.~\ref{fig:NV_location} (c) [(d)]; we optimized their geometries in large supercells (Fig.~S1) using DFT with the Perdew-Burke-Ernzerhof (PBE) functional~\cite{perdew_generalized_1996}, in either the singlet or triplet spin state, obtaining their relative formation energy ($\Delta E_{\rm f}$). Consistent with previous reports~\cite{ghassemizadeh_stability_2022}, we find that most of the NV centers have a lower formation energy ($E_{\rm f}$) than in the bulk [Figs.~\ref{fig:NV_location} (e-f), S2, and S3]. A substantial fraction is also stable in the triplet spin state. Among the 90 NVs near the 30$^{\circ}$ dislocation core, 62 (69\%) favor the triplet spin state over the singlet, with 46 (51\%) exhibiting a triplet state that is at least 0.1 eV lower in energy. Similarly, for the 112 NVs near the 90$^{\circ}$ dislocation core, 76 (68\%) favor the triplet spin state, with 57 (51\%) exhibiting a triplet state that is at least 0.1 eV lower in energy. Since not all centers considered here are stable in the desired triplet spin state, in a typical sample, some of the defects attracted to dislocations will be dark.

Among the screened configurations with low formation energies, we then selected a few representative NV defects (Fig.~S4) that are stable in a negative charge state and a triplet spin state, at least 0.1 eV lower in energy than the singlet state (to avoid possible temperature-induced transitions); their properties were examined in detail as discussed below.

All selected representative NVs are metastable compared to the most stable one, i.e., the G$^-$9-G11 (G$^-$6-G9) near the 30$^{\circ}$ (90$^{\circ}$) dislocation core; however, we expect that metastable configurations may be formed as a result of kinetic trapping. To understand the charge stability of these NVs, we computed their formation energy as a function of the system Fermi level ($E_{\rm F}$), and obtained the charge transition levels (CTL). We used the dielectric-dependent hybrid (DDH) functional~\cite{skone_selfconsistent_2014} and considered PBE optimized geometries. We find that all representative defects have lower $E_{\rm f}$ than in bulk diamond, for all possible $E_{\rm F}$ values within the gap (Fig.~S5). Despite the variations of CTL positions, all representative defects can be stable in charge states from +1 to $-2$ depending on the chosen value of $E_{\rm F}$. Importantly, they can be stabilized in the $-1$ charge state for a wide range of $E_{\rm F}$. An effective way to tune the Fermi level could be, for example, through the control of N dopants. We note that the dislocations studied in this work do not interfere with the tuning and controlling of $E_{\rm F}$ (Fig.~S5) as they do not introduce deep states in the band gap of the solid.

\begin{figure}[!ht]
\centering
\includegraphics[width=0.8\textwidth]{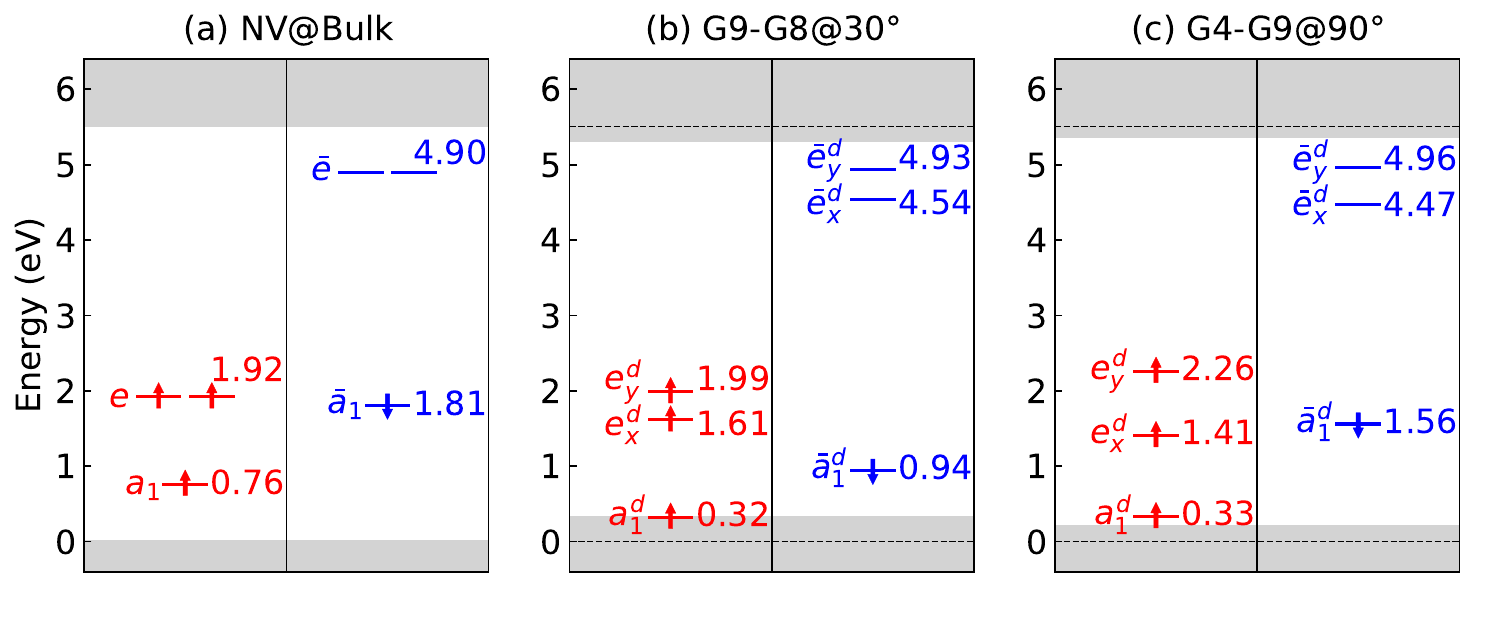}
\caption{
\textbf{DFT single-particle levels computed at the DDH level of theory.} (a) Bulk NV. (b) G9-G8 NV at the 30$^\circ$ dislocation core. (c) G4-G9 NV at the 90$^\circ$ dislocation core. See Fig.~\ref{fig:NV_location} and SI for the NV labeling.
Levels are shown in red (blue) for the spin-up (spin-down) channel, and next to each of them, we indicate the energy separation (in eV) relative to the valence band edge of diamond. The gray areas represent the valence and conduction bands of the system; the band edges of bulk diamond are indicated by the horizontal dashed lines in (b) and (c).
In (a), the defect levels are labeled as $a_1$ and $e$ ($\bar{a}_1$ and $\bar{e}$) in the spin-up (spin-down) channel according to the irreducible representations of the $C_{3v}$ group.
In (b) and (c), they are labeled as $a_1^d$, $e_x^d$, and $e_y^d$ ($\bar{a}_1^d$, $\bar{e}_x^d$, and $\bar{e}_y^d$) in the spin-up (spin-down) channel, with the superscript $d$ indicating the lowering of the $A_1$ and $E$ symmetries due to the presence of the dislocation. These labels are used here to illustrate the correspondence between the defects near dislocations and in the bulk.}
\label{fig:ground_state}
\end{figure}

Fig.~\ref{fig:ground_state} presents a comparison of the single-particle states of the bulk NV and the G9-G8 (G4-G9) NV at the 30$^{\circ}$ (90$^{\circ}$) dislocation. Results for other representative defects are presented in Figs.~S6 and S7. As expected, the symmetry of the defects is lowered close to dislocations, and thus the $e$ defect states, degenerate in bulk diamond, are split into two distinct levels [Fig.~\ref{fig:ground_state} (b) and (c)]. Not surprisingly, this splitting is observed for all representative NVs, and its magnitude, as well as the position of the defect levels within the gap, vary depending on the configuration.

Nevertheless, several important properties for the functionality of NV centers as qubits are shared by all configurations: the highest occupied ($\bar{a}_1^d$) and the lowest empty defect level ($\bar{e}_x^d$) in the spin-down channel are well separated from the electronic states of the host (Figs.~\ref{fig:ground_state}, S6, and S7). Further, the energy spacing between the $\bar{a}_1^d$ and $\bar{e}_x^d$ levels is sufficiently large ($>$2 eV) to prevent intra-defect thermal excitations near room temperature. Finally, the strain fields and distinct bonding geometries of dislocations give rise to electronic states within the gap of bulk diamond that are delocalized along the dislocation core and close in energy to the band edges of the solid. We find that these states are well separated from the defect levels (Figs.~\ref{fig:ground_state}, S6, and S7), as also shown by the analysis of localization factors reported in Table~S1.

\subsection{Excited-state properties of NV centers at dislocations}

We now turn to describe the excited-state properties of the representative NVs introduced above, computed using TDDFT with the DDH functional. These calculations employed recently developed methods and optimized GPU-accelerated codes~\cite{yu_gpu_2024,jin_excited_2023}. In particular, we computed and analyzed the intra-defect vertical excitation energies (VEEs) between many-body states. Fig.~\ref{fig:isc_rates} shows a comparison of the many-body state diagrams of the bulk NV and NV centers near dislocations; the diagrams for the other representative NVs are presented in Figs.~S8 and S9. We observe two triplet excited states [$^3A$(2) and $^3A$(3)] originating from spin-conserving excitations within the spin-down channel, where an electron is excited from $\bar{a}_1^d$ to one of the split $\bar{e}_x^d$ and $\bar{e}_y^d$ levels (Fig.~\ref{fig:ground_state}). The spin-flip excitations from $e_x^d$ and $e_y^d$ to $\bar{e}_x^d$ and $\bar{e}_y^d$ lead to transitions from $^3A$(1) to $^1A$(1), $^1A$(2) and $^1A$(3) states.

Similar to single-particle states, the many-body state diagrams of the NVs near the dislocation core vary depending on the configuration. However, for all representative spin defects, except for the G$^-$9-G$^-$10 near the 30$^{\circ}$ dislocation, the $^3A$(1) $\rightarrow$ $^3A$(2) transition is the lowest spin-conserving vertical excitation occurring in the spin-down channel, and the corresponding VEEs are smaller than those of defect-to-band transitions (Table~S2), similar to what was found for the bulk NV. Hence, our calculations show that a key requisite to prepare and measure the qubit state is satisfied for NV centers close to dislocations: the intra-defect optical transitions do not interfere with the electronic states of the host. Further, we observe three singlet excited states [$^1A$(1), $^1A$(2) and $^1A$(3)] positioned between the two triplet states $^3A$(1) and $^3A$(2) (Figs.~\ref{fig:isc_rates}, S8 and S9), indicating that an optical initialization cycle could, in principle, be realized that resembles that of the bulk NV. However, in some cases, e.g., the G3-G8 configuration near the core of 30$^{\circ}$ dislocation (Fig.~S8), the singlet excited state $^1A$(3) is higher in energy than the lowest triplet excited state $^3A$(2), and thus unlikely to be part of the non-spin-conserving decay path of the $^3A$(2) state. We also find that the VEEs from the ground triplet to the lowest singlet excited state for NVs near the dislocation core region are all smaller than in the bulk, suggesting an increased stability of singlet spin states near the core. Our calculations further show that the VEEs for the $^3A$(1) $\rightarrow$ $^1A$(1) transition are generally larger than the thermal energy at room temperature (0.025 eV), thus preventing thermal excitations; however, there are exceptions with a small VEE, e.g., the G10-G9 NV near the core of 90$^{\circ}$ dislocation, and these configurations are not expected to be useful qubits at least near room temperature.

We also computed the zero-phonon line (ZPL) energy for the spin defects in the bulk and at dislocations. Specifically, we computed the ZPL for the transition from the first triplet excited state to the ground state [$^3E$ $\rightarrow$ $^3A_2$ or $^3A$(2) $\rightarrow$ $^3A$(1); Figs.~\ref{fig:isc_rates}, S8, and S9]. Experimentally, ZPL values can be obtained from photoluminescence (PL) spectra. Hence, our results not only help characterize NVs near the core region but also assist in their experimental identification. Fig.~S10 summarizes the computed ZPL values and the Debye-Waller factors (DWF). Depending on the configuration and its proximity to the dislocation core, the computed ZPL can vary substantially (from 1.06 to 2.49 eV), compared to that (2.07 eV) in bulk diamond. Similar to the ZPL, NVs at dislocations exhibit varied DWFs. The largest (smallest) DWF obtained for the representative NVs is 10.17\% (0.55\%), compared to that (4.89\%) in bulk diamond, showing that all these defects may be observed by experiments with a detectable ZPL signal.

\subsection{Inter-system crossing rates and optical cycle}

\begin{figure}[!ht]
\centering
\includegraphics[width=0.8\textwidth]{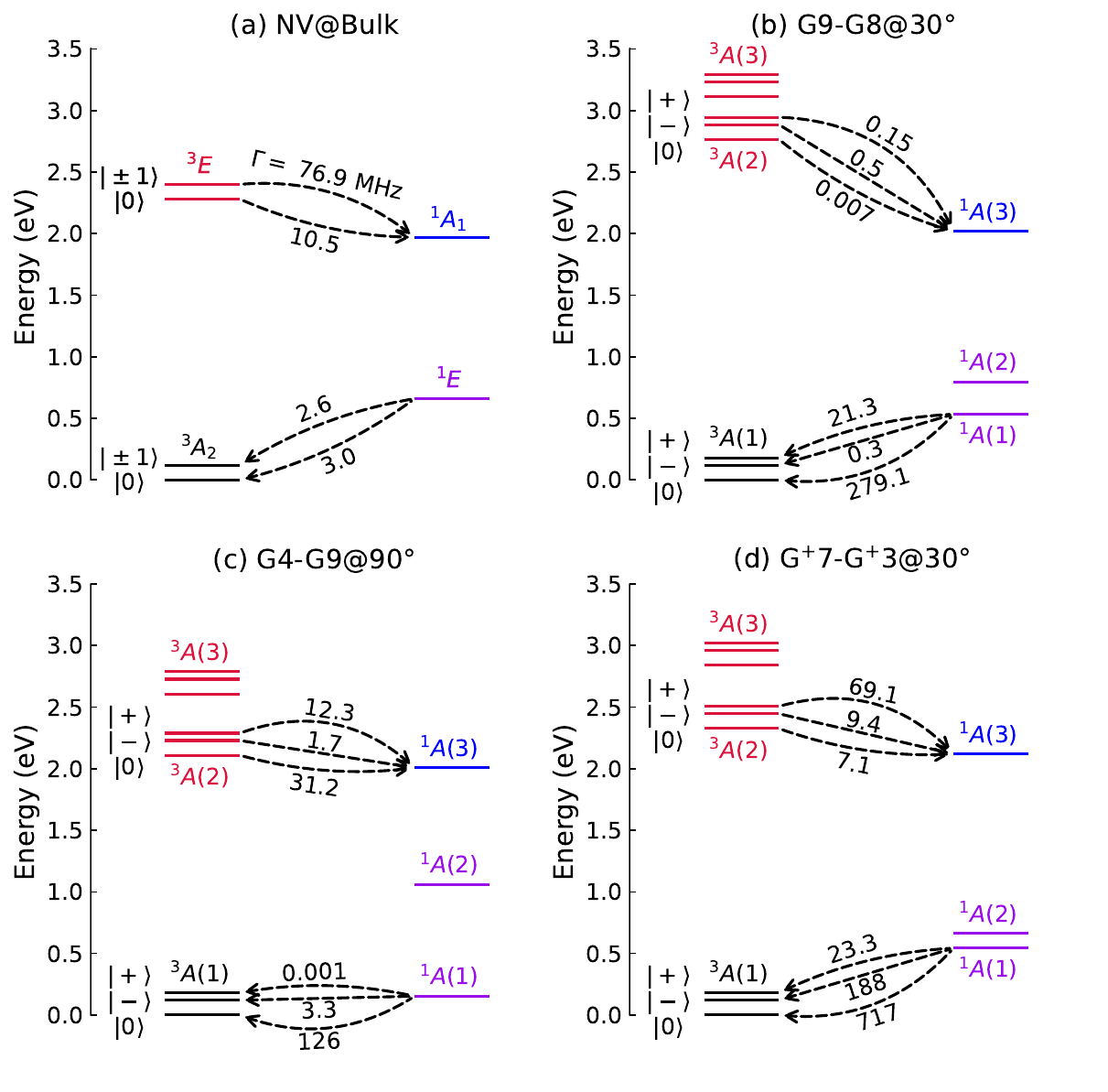}
\caption{
\textbf{Many-body electronic states diagram, derived from the vertical excitation energies (VEEs) obtained using TDDFT calculations at the DDH level of theory, and inter-system crossing (ISC) rates.}
(a) Bulk NV. (b) G9-G8 NV at the 30$^\circ$ dislocation core. (c) G4-G9 NV at the 90$^\circ$ dislocation core. (d) G$^+$7-G$^+$3 NV at the 30$^\circ$ dislocation core. See Fig.~\ref{fig:NV_location} and SI for the NV labeling.
In (a), the many-body triplet (singlet) states are labeled as $^3A_2$ and $^3E$ ($^1E$ and $^1A_1$) according to the irreducible representations of the $C_{3v}$ group.
In (b), (c), and (d), the many-body triplet (singlet) states are labeled as $^3A$(1), $^3A$(2), and $^3A$(3) [$^1A$(1), $^1A$(2) and $^1A$(3)] in order of increasing energy.
The ISC rate (in MHz) between two many-body states is shown next to the arrow connecting the corresponding states.}
\label{fig:isc_rates}
\end{figure}

To understand the optical spin initialization and readout processes of NVs near the dislocation core, we computed the ISC rates of three prototypical configurations; our results are shown in Fig.~\ref{fig:isc_rates}. As mentioned in the introduction, we employed here methodologies and codes that only recently became available to study ISC processes with advanced first-principles techniques~\cite{jin_first_2025}. We considered the experimental values for the bulk NV as reference~\cite{robledo_spin_2011}; we then computed the variation of the spin-orbit coupling (SOC) matrix elements between the triplet and singlet states using the many-body wavefunctions obtained from the quantum defect embedding theory (QDET)~\cite{sheng_green_2022,chen_advances_2025}, and the vibrational overlap functions for the bulk NV, at the respective estimated zero-phonon transition energies~\cite{jin_first_2025} (see SI). The symmetry lowering of the spin defects close to dislocations, from $C_{3v}$ to $C_{1}$, lifts the zero-field degeneracy of the $\ket{\pm 1}$ states, which hybridize to form the non-degenerate levels $\ket{\pm} = \frac{1}{\sqrt{2}}\Bigl(\ket{-1} \pm \ket{+1}\Bigr)$. Fig.~\ref{fig:isc_rates} shows the many-body electronic states diagrams for bulk NV and NVs near the core of dislocations. For simplicity, here we only consider the ISC transitions between $^3A(2)$ and $^1A(3)$ (denoted as upper ISC), and between $^1A(1)$ and $^3A(1)$ (lower ISC). The ISC transition rates from $^3A(3)$ to $^1A(3)$, and from $^1A(2)$ to $^3A(1)$ are assumed to be small compared to the non-radiative transition from $^3A(3)$ to $^3A(2)$ and from $^1A(2)$ to $^1A(1)$, given the small energy gap between these levels.

We recall that for the bulk NV, the upper ISC rates show a preferential selectivity between the $\ket{\pm 1}$ sub-levels of $^3E$ and $^1A_1$, which is key to its optical spin initialization and readout processes, while the lower ISC rates are much smaller and less selective. We find that the ISC rates of three NVs near the dislocation core analyzed here show quantitatively different behaviors, as illustrated in Fig.~\ref{fig:isc_rates} (b-d). For the G9-G8 configuration near the core of 30$^{\circ}$ dislocation [Fig.~\ref{fig:isc_rates} (b)], the upper ISC rate is much smaller than in bulk NV, due to the increased energy gap between $^3A(2)$ and $^1A(3)$; the selectivity for the ISC transitions from $\ket{+/-}$ sub-levels to $^1A(3)$ is similar to bulk NV; the lower ISC rates are faster than bulk NV, due to the decreased energy gap between $^1A(1)$ and $^3A(1)$, and the selectivity is higher for the $\ket{0}$ spin sub-level. For the G4-G9 configuration near the core of 90$^{\circ}$ dislocation [Fig.~\ref{fig:isc_rates} (c)], the upper ISC rates are comparable in magnitude to those in bulk NV, and they show an ISC selectivity from the $\ket{0}$ sub-level of $^3A(2)$ to $^1A(3)$; the lower ISC rate is also selective from the $^1A(1)$ to the $\ket{0}$ sub-level of $^3A(1)$ and faster than in bulk NV. For the G$^+$7-G$^+$3 configuration near the core of 30$^{\circ}$ dislocation [Fig.~\ref{fig:isc_rates} (d)], the upper ISC rates are comparable in both magnitude and selectivity to bulk NV; the lower ISC rates are much faster, given the decreased energy gap between $^3A(2)$ and $^1A(3)$ and increased SOC matrix elements, due to the breaking of the $C_{3v}$ symmetry. Based on the zero-phonon energy gaps and vibrational overlap functions of the 18 representative NVs (Fig.~S11), we estimate four of them (22\%) to exhibit desirable ISC rates.

\begin{figure}[!ht]
\centering
\includegraphics[width=0.8\textwidth]{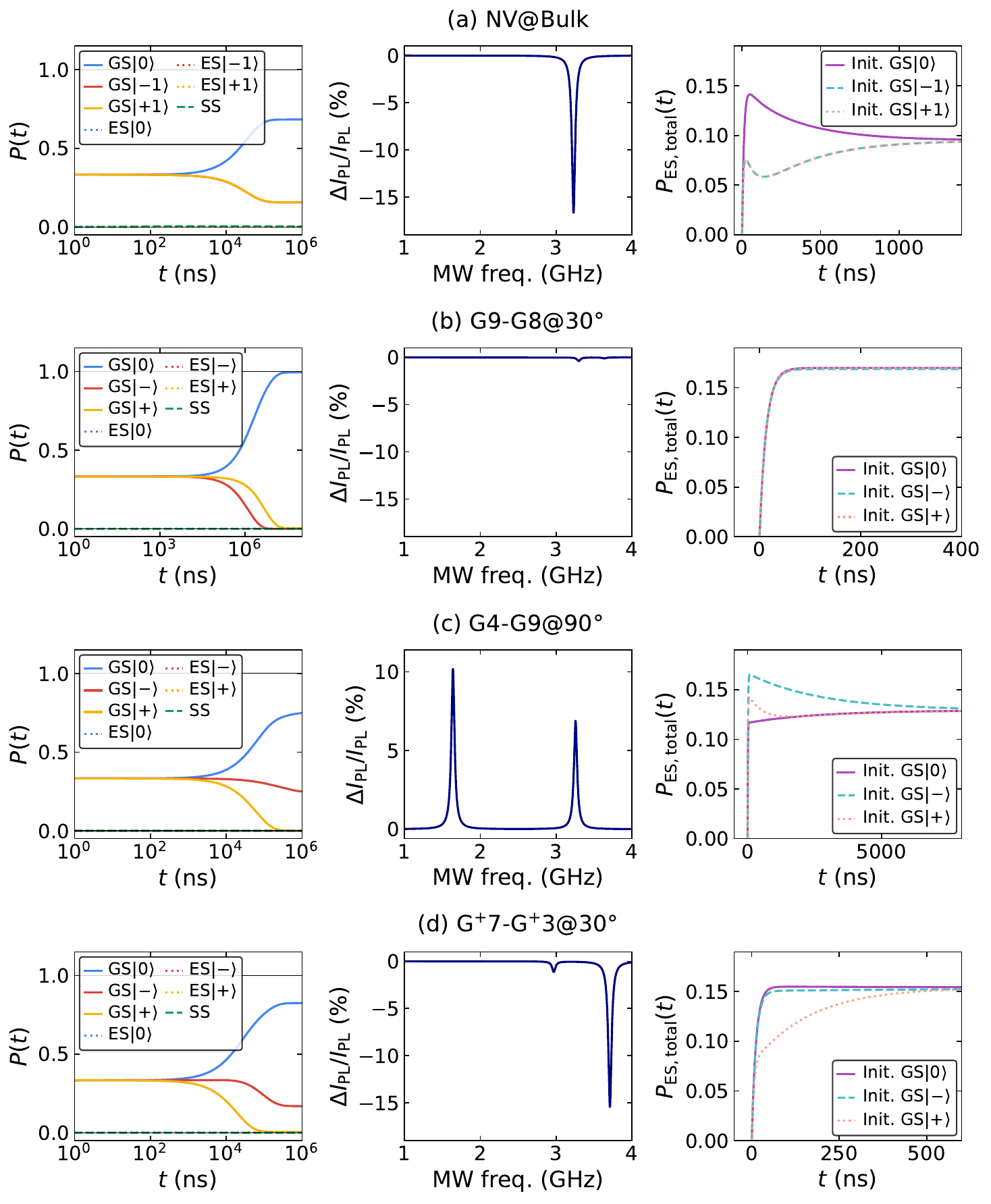}
\caption{
\textbf{State population simulations for the initialization, optically detected magnetic resonance (ODMR), and readout processes of NV qubits.}
(a) Bulk NV. (b) G9-G8 NV at the 30$^\circ$ dislocation core. (c) G4-G9 NV at the 90$^\circ$ dislocation core. (d) G$^+$7-G$^+$3 NV at the 30$^\circ$ dislocation core. See Fig.~\ref{fig:NV_location} and SI for the NV labeling.
The left panels display the change of the population of different spin sub-levels of the triplet ground state (GS) and excited state (ES) and the singlet state (SS) as a function of the time ($t$) of the optical initialization process; the initial population is set to 1/3 for each of the GS spin sub-levels. The middle panels display the continuous wave ODMR simulations; the left (right) peak corresponds to the GS $\ket{0}$ to GS $\ket{-}$ (GS $\ket{0}$ to GS $\ket{+}$) transition enabled by the microwave (MW).
The right panels display the photoluminescence (PL) spin readout simulations, where each line shows the change of the total population of the ES as a function of $t$, when the system is initialized into different sub-levels of the GS.}
\label{fig:odmr}
\end{figure}

Importantly, the essential properties required to enable an optical cycle are maintained for some NVs close to the core, as shown by state population simulations in Fig.~\ref{fig:odmr}. We considered a seven-state model (Fig.~S12): $\ket{+}$, $\ket{-}$, and $\ket{0}$ ($\ket{+1}$, $\ket{-1}$, and $\ket{0}$ for the bulk NV) spin sub-levels of the triplet ground state (GS), triplet excited state (ES), and one singlet state (SS); the latter is chosen to encompass the $^1A(1)$, $^1A(2)$, and $^1A(3)$ singlets ($^1E$ and $^1A_1$ singlets for the bulk NV) considering the relatively fast non-radiative transition rates between the various singlet states~\cite{jin_first_2025}. We performed three types of simulations for the optical spin initialization, the continuous wave optically detected magnetic resonance (cw-ODMR), and the PL spin readout (see SI for details).

From Fig.~\ref{fig:odmr} (a), we see that the bulk NV can be initialized into the $\ket{0}$ spin sub-level up to about 70\%, due to the selectivity of its upper ISC. Its ODMR contrast can reach about 18\% under the simulation conditions, showing its applicability in quantum sensing. Its PL spin readout shows a distinct difference for initialization into the $\ket{0}$ or $\ket{\pm 1}$ sub-levels, thus allowing for the readout of the population on different spin sub-levels using time-resolved PL measurements. These results are in agreement with what was previously reported for NVs in bulk diamond~\cite{ernst_modeling_2023,ernst_temperature_2023,li_excited_2024}.

For the G9-G8 configuration near the core of 30$^{\circ}$ dislocation [Fig.~\ref{fig:odmr} (b)], the initialization takes $\simeq$ 100 times longer to reach the steady state than in bulk NV, due to the much smaller upper ISC rates, and the ODMR contrasts are 40 times smaller; its time-resolved PL displays negligible differences between different initialization conditions, implying that this NV may not be useful for quantum applications based on optical control. The G4-G9 configuration near the core of 90$^{\circ}$ dislocation [Fig.~\ref{fig:odmr} (c)] can be initialized to the $\ket{0}$ sub-level up to about 75\% because of the much faster lower ISC to the GS $\ket{0}$ sub-level; its ODMR has a positive contrast, given the reversed selectivity of the upper ISC, compared to that of the bulk NV; its time-resolved PL exhibits a noticeable difference between initialization into the $\ket{0}$ or $\ket{-}$ sub-level. The G$^+$7-G$^+$3 configuration near the core of 30$^{\circ}$ dislocation [Fig.~\ref{fig:odmr} (d)] exhibits an initialization time similar to the bulk NV; it has a small ODMR contrast between the $\ket{0}$ and $\ket{-}$ sub-levels because of the comparable triplet ES $\ket{-}$ to SS and triplet ES $\ket{0}$ to SS ISC rates; its time-resolved PL displays a noticeable difference between the $\ket{0}$ and $\ket{+}$ sub-levels and can be potentially used for PL spin readout.

The state population simulations show that there exist NV configurations close to dislocations that can be properly initialized [Fig.~\ref{fig:odmr} (c) and (d)] and used in an optical cycle similar to that of the bulk. In addition, our results provide a means to experimentally identify configurations with specific characteristics close to dislocations.

\subsection{Coherence times}

\begin{figure}[!ht]
\centering
\includegraphics[width=0.8\textwidth]{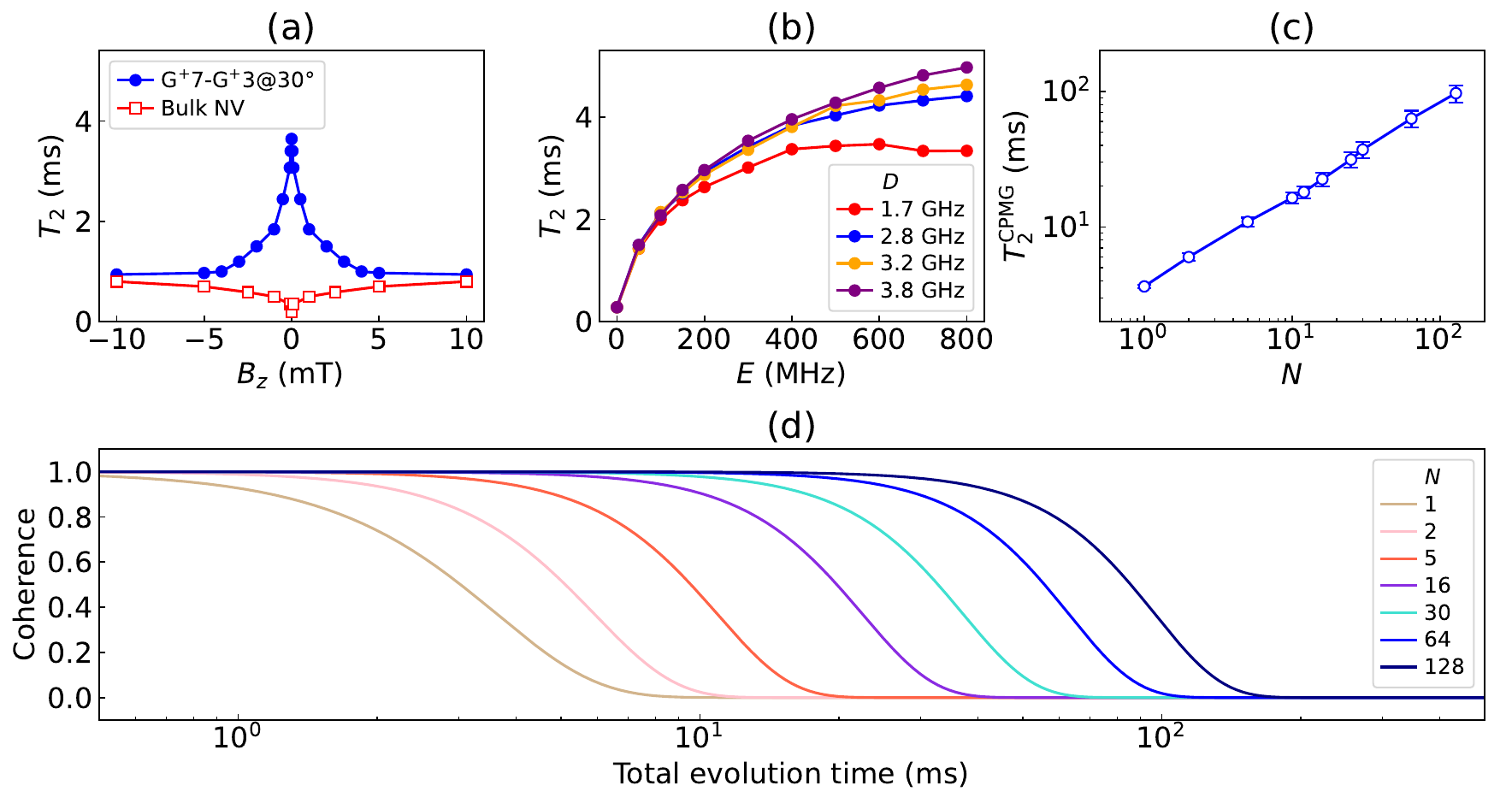}
\caption{
\textbf{Coherence times of NV centers near dislocations.}
(a) Clock transitions at weak magnetic field for the G$^+$7-G$^+$3 NV near the core of 30$^{\circ}$ dislocation. (b) Computed $T_2$ times at zero magnetic field as a function of the transverse zero-field splitting (ZFS) $E$, considering different axial $D$ values. (c) Computed $T^{\rm CPMG}_2$ times at zero magnetic field at different decoupling pulses $N$ for the G$^+$7-G$^+$3 NV near the core of 30$^{\circ}$ dislocation. (d) Coherence functions of the G$^+$7-G$^+$3 NV near the core of 30$^{\circ}$ dislocation at zero magnetic field for various decoupling pulses $N$.}
\label{fig:CT}
\end{figure}

Finally, we discuss coherence times of representative NV defects close to dislocations, and we start by evaluating their axial ($D$) and transverse ($E$) components of the ground-state zero-field splitting (ZFS) tensor (Fig.~S13). The highest (lowest) $D$ value is found to be 3839.0 (1796.6) MHz, deviating by up to 44\% from that of the bulk NV (3230.8 MHz). The lowering of the defect symmetry leads to large nonzero $E$ values for the defects close to dislocations, with the highest predicted value reaching 813.3 MHz. Hence, the zero‐field degeneracy of the $\ket{\pm1}$ states is lifted to form non‐degenerate ``clock'' levels $\ket{\pm}$ with no first‐order magnetic moment. As a result, NV centers at dislocations exhibit first‐order‐insensitive transitions in weak magnetic fields ($B$), known as ``clock transitions''. We computed the Hahn‐echo coherence time $T_2$ for magnetic fields up to $10$ mT [Figs.~\ref{fig:CT} (a), S14, and S15] and observe that $T_2$ decreases with increasing $B$, reflecting enhanced decoherence from the field‐dependent dynamics of the $^{13}$C spin bath as one moves away from the zero‐field clock transition. This behavior is notably different from that of bulk NV (whose $E$ component of the ZFS is zero), which monotonically increases with $B$ and saturates at $\sim$1 ms [Fig~\ref{fig:CT} (a)]~\cite{zhao_decoherence_2012}.

The representative NV centers near dislocations exhibit longer coherence times than the bulk NV, at 0 mT up to $\sim$1 mT depending on the ZFS parameters (Figs.~S14 and S15). The observed enhancement of $T_2$ at $B=0$ is determined by both the $D$ and $E$ components of the ZFS tensor. As illustrated in Fig.~\ref{fig:CT} (b), an increase in $D$ and $E$ leads to a significant enhancement in $T_2$, reaching up to an order of magnitude improvement compared to bulk NV. A similar enhancement is also observed in the free induction decay time $T^*_2$ (Fig.~S16). To further investigate the protection of spin coherence from environmental noise, we simulated NV coherence as a function of total evolution time under periodic Carr-Purcell-Meiboom-Gill (CPMG) pulse sequences at $B=0$ on representative NVs near dislocations. As shown in Fig.~\ref{fig:CT} (c), the coherence time of the $G^+7$-$G^+3$ NV at the $30^\circ$ dislocation increases from 3.7 ms to $\sim$100 ms with $N=128$ $\pi$-pulses, and the traces in Fig.~\ref{fig:CT} (d) show that raising $N$ suppresses decoherence and steepens the decay. We note that the reported $T^{\rm CPMG}_2$ values were computed in the long-$T_1$ limit, where longitudinal relaxation does not affect NV coherence~\cite{balasubramanian_ultralong_2009,abobeih_one_2018}. Our results demonstrate that the ZFS tensor of the NV can be engineered via dislocations, significantly enhancing coherence time and providing optimal working points for NV qubits.

\section{Discussion}

We investigated the feasibility of using NV centers as spin qubits close to dislocations in diamond. We conducted a high-throughput screening of all possible NV configurations near the core regions of two types of dislocations. Our formation energy calculations indicate that NVs are energetically attracted to dislocation cores, suggesting that dislocations can serve as a scaffold to position NVs in a 1D arrangement. We further showed that multiple NV configurations near the core can be stabilized in a triplet spin state and $-1$ charge state. We then selected representative NVs and thoroughly examined their single-particle defect levels, optical cycle, and coherence times. In spite of differences in these properties relative to the bulk NVs, we found representative spin defects in the core region that meet the requisites to be useful qubits. Importantly, the optical cycle for qubit initialization and readout is feasible for several NVs. Moreover, certain centers can outperform the bulk NV, exhibiting a larger DWF and improved coherence time.

We emphasize that the optical cycle of spin defects close to dislocations has never been investigated before; we could address this problem here thanks to the integration of a comprehensive suite of accurate and scalable simulation techniques that have only recently been developed~\cite{jin_first_2025,yu_gpu_2024,jin_excited_2023,chen_advances_2025}, enabling the investigation of the interplay between extended and point defects. We also note that given the large system size required to account for both point and extended defects (we used a 1728-atom supercell of diamond), our massively parallel GPU-accelerated implementation of TDDFT and QDET in the WEST code~\cite{jin_excited_2023,chen_advances_2025,yu_gpu_2022,gavini_roadmap_2023} turned out to be critical to obtain accurate results, as was our recently developed workflow~\cite{jin_first_2025} to predict ISC rates.

Based on the promising properties identified in our study, specifically the favorable energetics and the preserved or enhanced qubit properties, we propose that NVs close to dislocations can be utilized to pattern and couple 1D arrays of spin defects, forming the conceptual basis for dislocation-based quantum interconnects. This engineering strategy to create quantum interconnects may advance the development of solid-state quantum technologies. Moreover, the approach is not limited to NV qubits in diamond and could be applied to other hosts such as SiC and other geometries, including semiconductor/oxide interfaces~\cite{son_developing_2020,zhang_optical_2024}.

However, clearly, some challenges remain to be addressed to fulfill the potential of the engineering strategy proposed here. Only a fraction of spin defects close to dislocations display the desired quantum properties. Hence, a controlled synthesis and manipulation of spin defects with favorable functionalities needs to be developed. Part of the synthetic strategy may include the controlled introduction and patterning of specific dislocations into the sample, as well as the use of impurities (e.g., N) to engineer the desired Fermi level values.

Recently, experimental efforts have increasingly explored NV centers in low-symmetry environments as a means to tune and engineer their properties. Quasi-1D arrays of NVs have been created along heavy-ion tracks in diamond~\cite{liu_optical_2025}, and NVs formed within diamond nanopillars have been demonstrated~\cite{kim_scalable_2025}. Additionally, NVs near surfaces~\cite{hughes_strongly_2025} and interfaces~\cite{lo_enhancement_2025} have been reported. The workflow established here can be readily extended to these systems, providing valuable insights to help interpret and guide experimental investigations in this regime.

\section{Methods}

\subsection{Details of first-principles calculations}

Using the 1728-atom supercells, we carried out a high-throughput search of configurations of NV centers close to dislocations using DFT as implemented in the Quantum ESPRESSO code~\cite{giannozzi_quantum_2020}. We then computed their electronic and coherence properties, and compared them with those of NVs in bulk diamond. Using DFT, we computed the structural properties of spin defects with the PBE~\cite{perdew_generalized_1996} functional, and formation energies in varied charge states, charge transition levels, and single-particle energies with the DDH functional~\cite{skone_selfconsistent_2014}. We computed many-body energy levels using (spin-flip) TDDFT with the DDH functional, as implemented in the WEST code~\cite{jin_excited_2023}. ISC rates were obtained with a newly developed framework~\cite{jin_first_2025} that combines calculations of many-body states with QDET~\cite{sheng_green_2022,chen_advances_2025}, fully relativistic calculations of SOC, and calculations of overlap functions using first-principles phonons~\cite{jin_photoluminescence_2021,jin_vibrationally_2022}. We used a seven-state model for state population simulations of ODMR spectra. Coherence times were obtained with the generalized cluster expansion techniques (gCCE)~\cite{yang_quantum_2008,onizhuk_probing_2021} and the PyCCE code~\cite{onizhuk_pycce_2021}, and calculations of ZFS parameters were carried out with the PyZFS code~\cite{ma_pyzfs_2020}. Details are provided in the SI.

\section{Data availability}
Data that support the findings of this study are openly available at \url{https://paperstack.uchicago.edu}.

\section{Code availability}
Codes are openly available in public repositories.

\begin{itemize}
    \item Quantum ESPRESSO: \url{https://gitlab.com/QEF/q-e}
    \item WEST: \url{https://github.com/west-code-development/West}
    \item PyCCE: \url{https://github.com/MICCoMpy/PyCCE}
    \item PyZFS: \url{https://github.com/MICCoMpy/pyzfs}
\end{itemize}

\section{Acknowledgements}
We thank F. Joseph Heremans, Alex High, Benchen Huang, and Marquis M. McMillan for useful discussions.
This research used resources of the National Energy Research Scientific Computing Center (NERSC), a U.S. Department of Energy Office of Science User Facility operated under Contract No. DE-AC02-05CH11231.
This research also used resources of the Argonne Leadership Computing Facility at Argonne National Laboratory, a U.S. Department of Energy Office of Science User Facility operated under Contract No. DE-AC02-06CH11357.
This work was supported by the AFOSR Grant No. FA9550-23-1-0330.
Software development (WEST, PyCCE, and PyZFS) in this work was supported by MICCoM, as part of the Computational Materials Sciences Program funded by the U.S. Department of Energy, Office of Science, Basic Energy Sciences, Materials Sciences and Engineering Division through Argonne National Laboratory.

\section{Author contributions}
M.G. and G.G. conceived the research ideas. C.Z., V.W.-z.Y., Y.J., J.N., and S.P.G. performed the calculations. All authors contributed to the data analysis and manuscript writing.

\section{Competing interests}

The authors declare no competing financial or non-financial interests.

\bibliography{disloc}

\section{Supplementary Information}
Supplementary Text, Figs. S1--S16, Tables S1--S4 and References.

\end{document}


\title{Supplementary information: ``Towards dislocation-driven quantum interconnects''}

\author[1]{\fontsize{14}{14}\selectfont Cunzhi Zhang \thanks{equally contributing author}}

\author[2]{\fontsize{14}{14}\selectfont Victor Wen-zhe Yu \thanks{equally contributing author}}

\author[1]{\fontsize{14}{14}\selectfont Yu Jin}

\author[1]{\fontsize{14}{14}\selectfont Jonah Nagura}

\author[3]{\fontsize{14}{14}\selectfont Sevim Polat Genlik}

\author[3]{\fontsize{14}{14}\selectfont Maryam Ghazisaeidi \thanks{Corresponding author: ghazisaeidi.1@osu.edu}}

\author[1,2,4]{\fontsize{14}{14}\selectfont Giulia Galli \thanks{Corresponding author: gagalli@uchicago.edu}}

\affil[1]{\fontsize{12}{12}\selectfont Pritzker School of Molecular Engineering, University of Chicago, Chicago, IL 60637, USA}
\affil[2]{\fontsize{12}{12}\selectfont Materials Science Division, Argonne National Laboratory, Lemont, IL 60439, USA}
\affil[3]{\fontsize{12}{12}\selectfont Department of Materials Science and Engineering, The Ohio State University, Columbus, OH 43210, USA}
\affil[4]{\fontsize{12}{12}\selectfont Department of Chemistry, University of Chicago, Chicago, IL 60637, USA}

\date{}

\maketitle

\section{Supercell used to simulate dislocations}

In cubic diamond, the primary slip system is \{111\}⟨110⟩. Dislocations are classified as glide or shuffle, depending on whether slip occurs between closely spaced or widely spaced \{111\} planes, respectively. In this study, we focus on the glide set of dislocations, as they are the most energetically favorable~\cite{pizzagalli_dislocation_2008}. Prior studies have demonstrated that full (60$^{\circ}$) glide dislocations lower their energy by dissociating into 30$^{\circ}$ and 90$^{\circ}$ Shockley partials on the \{111\} plane, thereby bounding a stacking fault~\cite{blumenau_dislocations_2002,blumenau_dislocations_2003}. These partial dislocations with edge and mixed characters are of great importance due to their strain fields, which create a long-range pressure field that interacts with point defects such as substitutional nitrogen (N) and vacancies, unlike the shear-dominated fields of pure screw dislocations. Edge dislocations, characterized by an extra half-plane of atoms, produce a pressure field that is compressive on the side of the glide plane where the extra half-plane is located and tensile on the opposite side. This configuration reduces the formation energy of vacancies and solutes with a negative misfit, such as N in diamond, on the compressive side, creating a thermodynamic driving force for their segregation to dislocations.

\begin{figure}[!ht]
\centering
\includegraphics[width=0.7\textwidth]{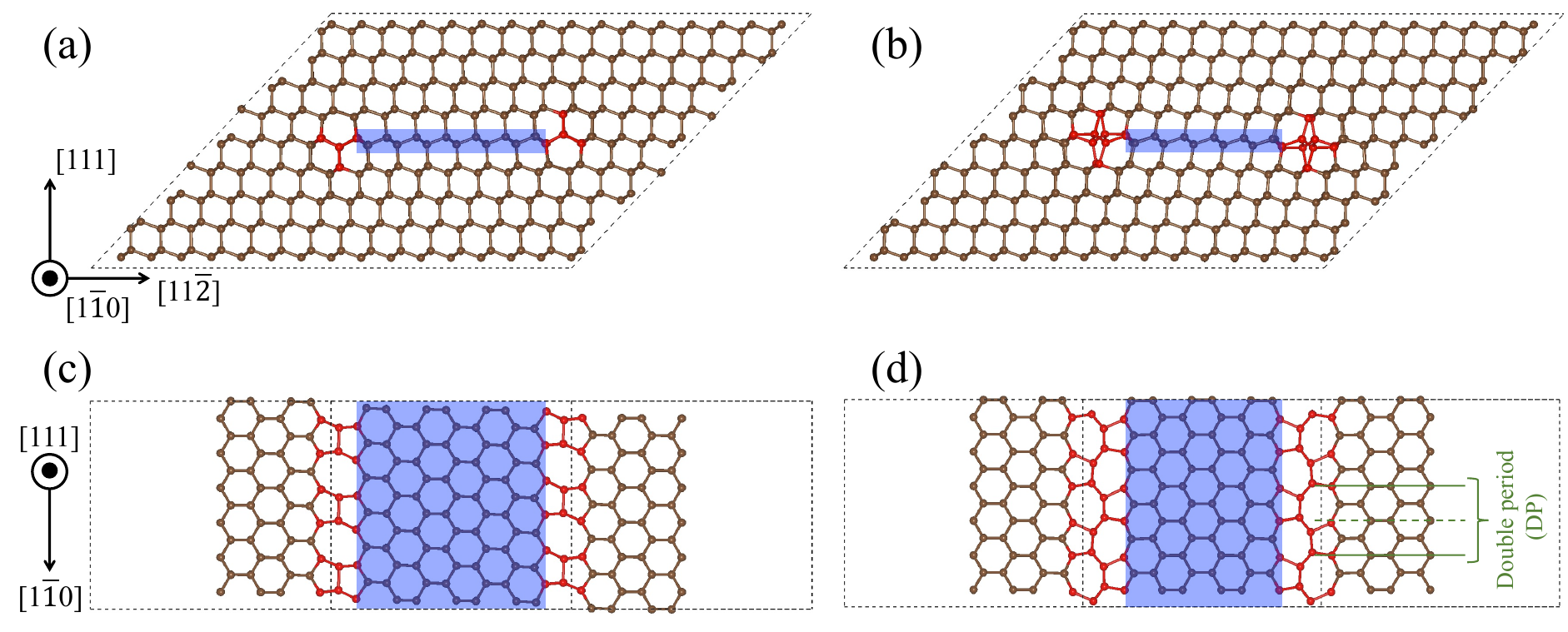}
\caption{
\textbf{Structure of the diamond supercell including two dislocations of opposite Burgers vectors.}
Atoms within the dislocation cores are shown in red; the stacking fault between the two partial dislocations is shaded in blue. As indicated in (d), the lattice translation period along the [1$\overline{1}$0] dislocation line direction is doubled compared to that of pristine diamond (the distance between the two solid green lines compared to that between the solid and dashed green lines) to allow for core reconstruction and removal of the dangling bonds.
(a) Supercell containing two 30$^{\circ}$ dislocations.
(b) Supercell containing two 90$^{\circ}$ dislocations.
(c) The glide plane containing the stacking fault between the two 30$^{\circ}$ dislocations.
(d) The glide plane containing the stacking fault between the two 90$^{\circ}$ dislocations.}
\label{s-fig:supercell}
\end{figure}

Isolated dislocations have long-range elastic fields that break the translational invariance of the crystal. To accurately model dislocations with full periodic boundary conditions, we use a quadrupolar arrangement in a monoclinic simulation cell with two opposite Burgers vector dislocations to cancel long-range elastic fields and minimize strain at the edges of the simulation cell. The supercell is oriented such that the x, y, and z directions correspond to [$\overline{1}\overline{1}$2], [111], and [1$\overline{1}$0] crystallographic directions, respectively. The dislocations are introduced by displacing all atoms according to their anisotropic elastic displacement field. During dislocation core geometry optimization, without point defects, the lattice translation period along the dislocation line direction (z or [1$\overline{1}$0]) is doubled to allow for core reconstruction. In calculations involving negatively charged nitrogen-vacancies (NVs), the supercell size along the dislocation line direction is chosen to be 15.1 \AA~to minimize the interaction between an NV defect and its periodic images. The simulation cell (without NV) consists of 1728 atoms (35.0 \AA $\times$ 25.5 \AA $\times$ 15.1 \AA), with a separation of $\sim$20 \AA~between the two dislocations, as shown in Fig.~\ref{s-fig:supercell}. We also constructed a diamond supercell of the same size but without dislocations, within which we created an NV defect to represent the NV defect in bulk diamond.

\section{Selection of NV configurations near the dislocation core}

We examined various NV configurations near the core. To create an NV defect, we first selected a pair of two adjacent carbon (C) atoms in the supercell, then substituted one of the C atoms with an N atom and removed the other C atom to generate a C vacancy (V$_{\rm C}$). Specifically, the positions of the N atom and the neighboring V$_{\rm C}$ for an NV defect are chosen within the G, G$^+$ and G$^-$ planes [see Fig.~1 (a-b)]; the corresponding atomic sites near the 30$^{\circ}$ (90$^{\circ}$) dislocation core are numbered as shown in Fig.~1 (c) [(d)]. We took into account the periodicity along the dislocation line and enumerated all geometrically distinct NV defects along the line direction. NV defects further away from the core beyond the orange, red, or blue atomic sites in Fig.~1 are not studied in this work, but will be considered in the future. We label NV defects based on their locations; for example, G12-G$^+$7 indicates that the N atom occupies position G12 and the V$_{\rm C}$ resides at G$^+$7.

\section{Formation energies of NV configurations near the dislocation core}

We employed density-functional theory (DFT) calculations using the plane-wave pseudopotential method as implemented in the Quantum ESPRESSO code~\cite{giannozzi_quantum_2020} to investigate NV configurations near the dislocation core, and all calculations were spin-polarized. The SG15 optimized norm-conserving Vanderbilt (ONCV) pseudopotentials~\cite{hamann_optimized_2013,schlipf_optimization_2015} were used. The exchange-correlation was treated at two levels of theory, namely the semilocal Perdew-Burke-Ernzerhof (PBE)~\cite{perdew_generalized_1996} functional and the dielectric dependent hybrid (DDH) functional~\cite{skone_selfconsistent_2014}. The latter includes a fraction of the exact exchange with the mixing parameter $\alpha = 0.18$ set equal to the inverse of the macroscopic dielectric constant of bulk diamond, assuming that the presence of the dislocation would not drastically change the macroscopic dielectric constant of the material. The kinetic energy cutoff for the plane-wave basis set was set to 60 Ry for the wavefunctions and 240 Ry for the electron density. The Brillouin zone of the supercell was sampled at the $\Gamma$-point. The computed band gap of bulk diamond, including zero-point renormalization~\cite{yang_computational_2022}, is 5.09 eV, in reasonable agreement with the experimental value of 5.48 eV~\cite{noauthor_intrinsic_1964}. The ground-state geometries of NV defects were optimized using DFT at the PBE level of theory. The atomic positions were relaxed until the forces acting on the atoms became smaller than $10^{-3}$ Ry/Bohr and the total energy changed less than $10^{-4}$ Ry between two consecutive relaxation steps. The spin state of an NV defect was controlled by constraining the number of electrons in the spin-up and down channels.

\begin{figure}[!ht]
\centering
\includegraphics[width=0.7\textwidth]{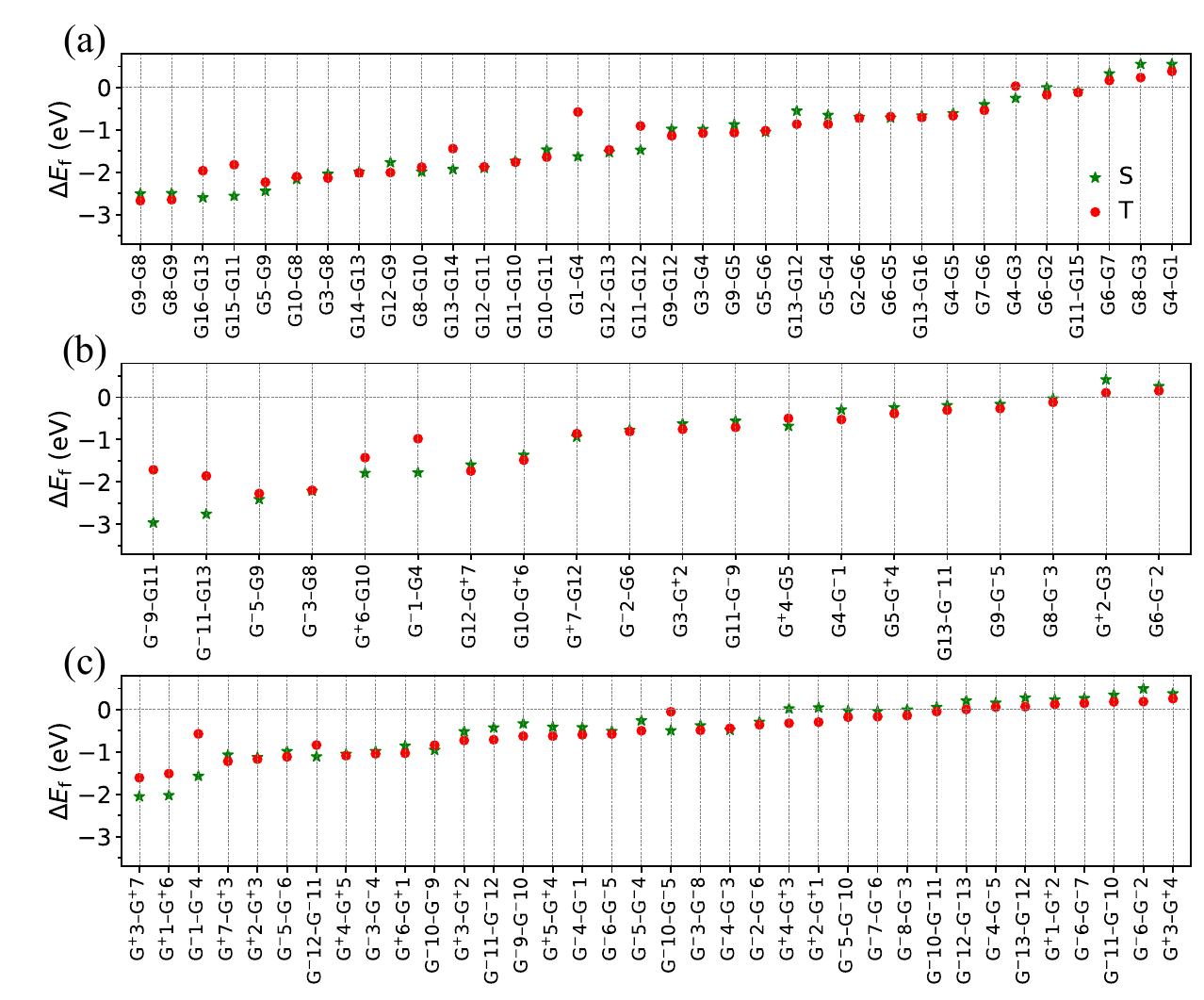}
\caption{
\textbf{Relative formation energy ($\Delta E_{\rm f}$) of NV near the 30$^{\circ}$ dislocation core with respect to that in bulk diamond.}
These results were obtained for NV in the $-1$ charge state at the PBE level of theory; S (T) stands for the computed results for relaxed NV defects in the singlet (triplet) spin state.
(a) NV defects located in the \{111\} glide plane (G).
(b) NV defects with either the nitrogen atom or the nearby carbon vacancy located in the \{111\} glide plane (G) and the other one of the two in an adjacent \{111\} plane (G$^+$ or G$^-$).
(c) NV defects with both the nitrogen atom and the nearby carbon vacancy located in the adjacent \{111\} planes (G$^+$ or G$^-$).
See Fig.~1 for the definition of the \{111\} glide planes and NV notations.}
\label{s-fig:Ef_30}
\end{figure}

\begin{figure}[!ht]
\centering
\includegraphics[width=0.7\textwidth]{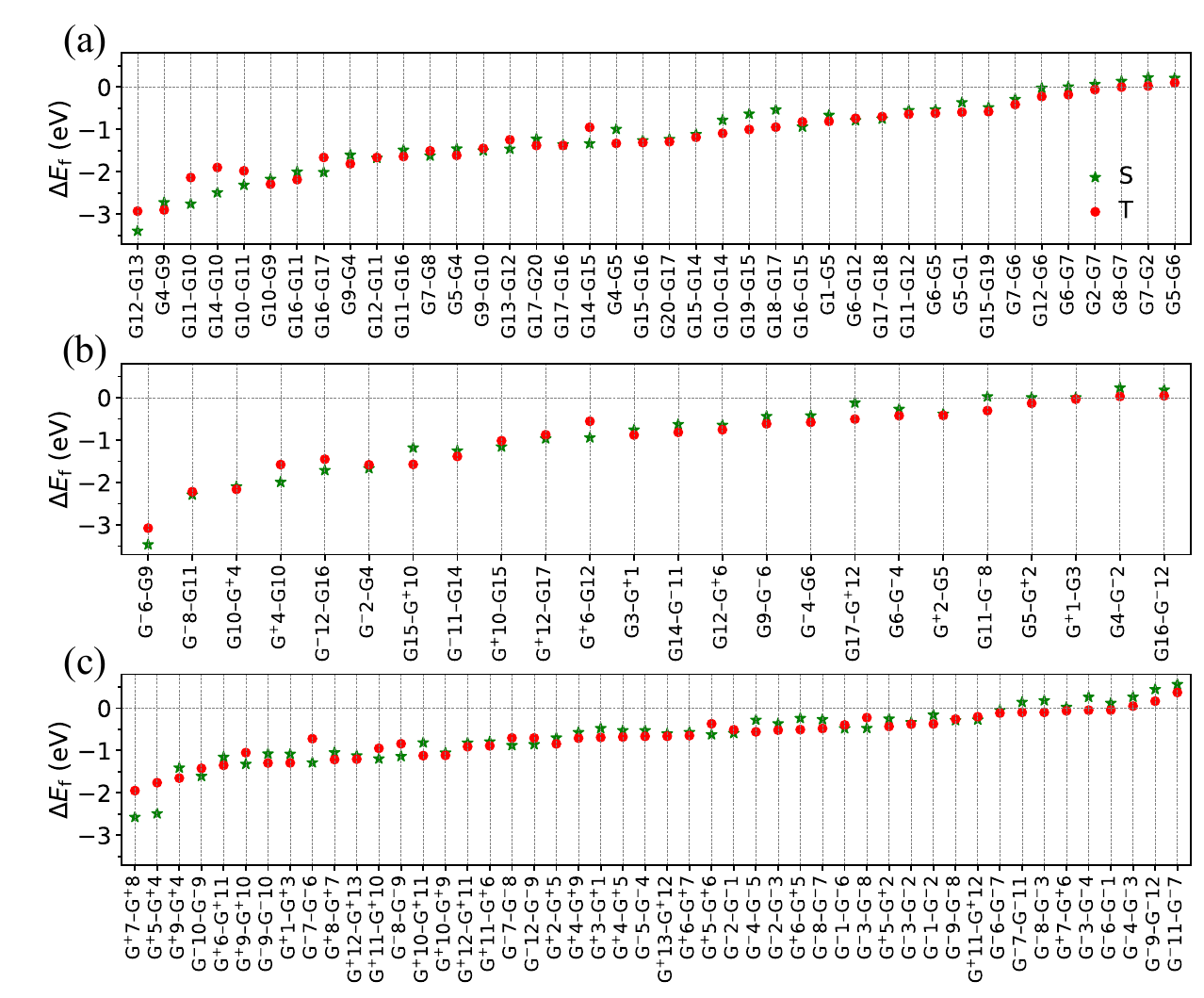}
\caption{
\textbf{Relative formation energy ($\Delta E_{\rm f}$) of NV near the 90$^{\circ}$ dislocation core with respect to that in bulk diamond.}
These results were obtained for NV in the $-1$ charge state at the PBE level of theory; S (T) stands for the computed results for relaxed NV defects in the singlet (triplet) spin state.
(a) NV defects located in the \{111\} glide plane (G).
(b) NV defects with either the nitrogen atom or the nearby carbon vacancy located in the \{111\} glide plane (G) and the other one of the two in an adjacent \{111\} plane (G$^+$ or G$^-$).
(c) NV defects with both the nitrogen atom and the nearby carbon vacancy located in the adjacent \{111\} planes (G$^+$ or G$^-$).
See Fig.~1 for the definition of the \{111\} glide planes and NV notations.}
\label{s-fig:Ef_90}
\end{figure}

We obtained the relative formation energy ($\Delta E_{\rm f}$) of NVs near the dislocation core in the $-1$ charge state with respect to that in bulk diamond as:
\begin{equation}
\begin{split}
\Delta E_{\rm f}({\rm NV^-@disloc}, s) =
\Big( E_{\rm tot}({\rm NV^-@disloc}, s) - E_{\rm tot}({\rm disloc}) \Big) \\ -
\Big( E_{\rm tot}({\rm NV^-@bulk}) - E_{\rm tot}({\rm bulk}) \Big)
\end{split}
\end{equation}
where ${\rm NV^-@disloc}$ stands for an NV defect in the $-1$ charge state (NV$^-$) near the dislocation core; ${\rm NV^-@bulk}$ stands for the bulk NV$^-$ center; $s$ denotes spin state, being either singlet (S) or triplet (T); $E_{\rm tot}( {\rm NV^-@disloc}, s)$ is the total energy of the 1727-atom supercell containing two dislocations with a relaxed NV defect near the dislocation core in spin state $s$; $E_{\rm tot}({\rm NV^-@bulk})$ is the total energy of the 1727-atom supercell without dislocations containing a relaxed NV defect in the triplet spin state; $E_{\rm tot}({\rm disloc})$ is the total energy of the 1728-atom supercell with two dislocations without any NV defect; and $E_{\rm tot}({\rm bulk})$ is the total energy of the 1728-atom supercell without dislocations and without an NV defect. All total energies were obtained at the PBE level of theory. These $\Delta E_{\rm f}$ results are shown in Figs.~\ref{s-fig:Ef_30} and \ref{s-fig:Ef_90}.

\section{Formation energies of representative NV configurations}

\begin{figure}[!ht]
\centering
\includegraphics[width=0.7\textwidth]{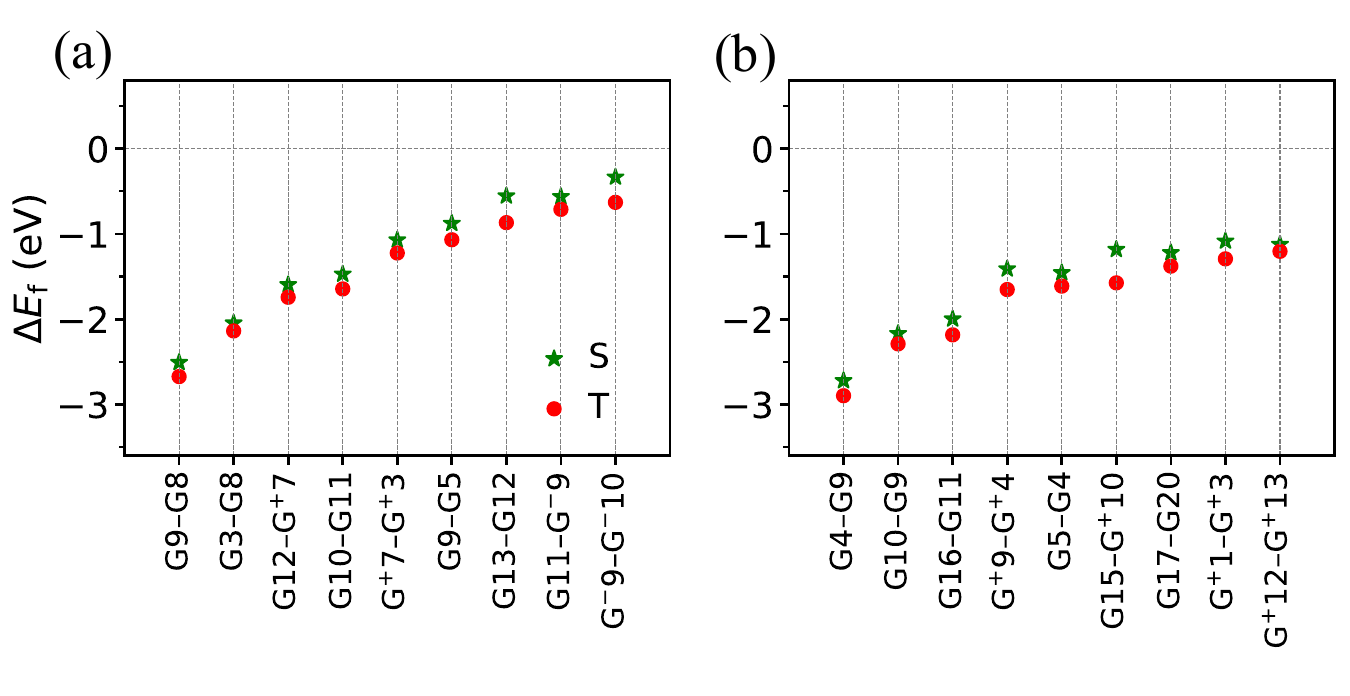}
\caption{
\textbf{Relative formation energy ($\Delta E_{\rm f}$) of representative NV defects near the dislocation core with respect to that in bulk diamond.}
These results were obtained for NV in the $-1$ charge state at the PBE level of theory; S (T) stands for the computed results for relaxed NV defects in the singlet (triplet) spin state. In the main text, we discuss how we selected these representative NV configurations.
(a) 9 representative NV defects near the 30$^{\circ}$ dislocation core.
(b) 9 representative NV defects near the 90$^{\circ}$ dislocation core.
See Fig.~1 for the NV notations.}
\label{s-fig:Ef_representative}
\end{figure}

For representative spin defects near the core region in the 1728-atom supercell (see the main text and Fig.~\ref{s-fig:Ef_representative}), we relaxed their geometries in different charge and spin states using DFT at the PBE level of theory. We considered charge state $q$ from $-3$ to 2, and spin state $s$ = S or T (D or Q) for even (odd) number of electrons in the system, where D stands for doublet and Q for quartet.

\begin{figure}[!ht]
\centering
\includegraphics[width=0.7\textwidth]{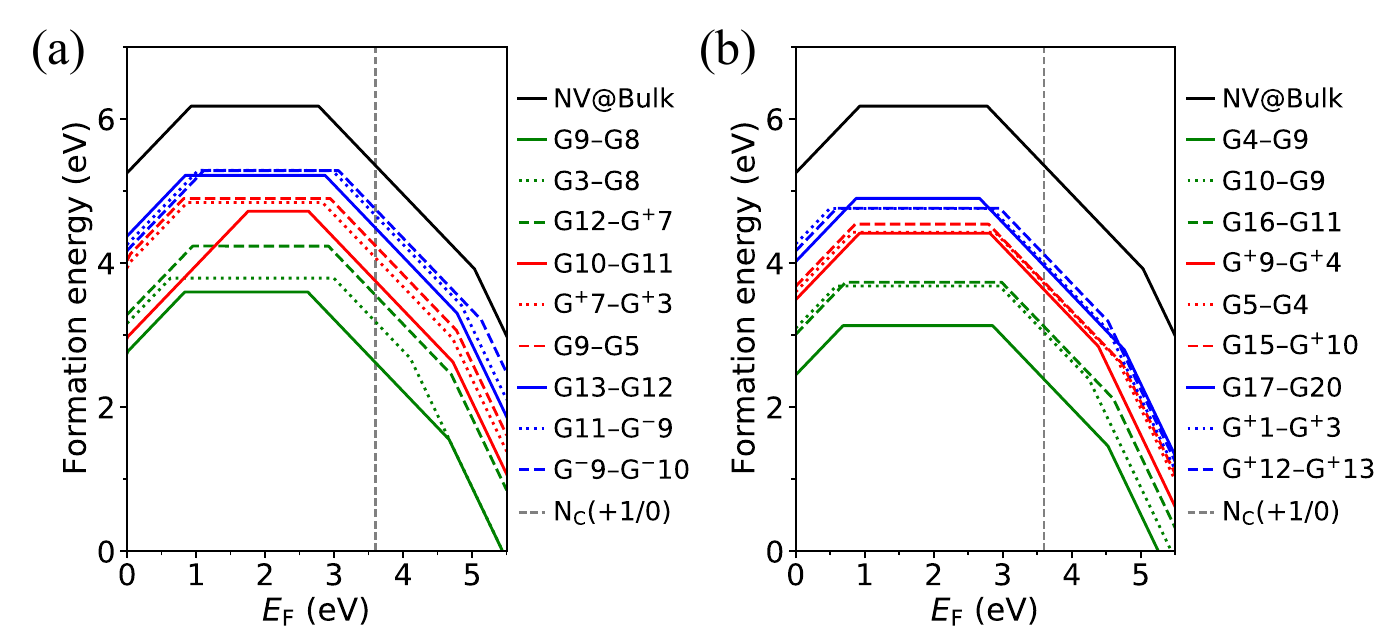}
\caption{
\textbf{Formation energy of the representative NV configurations as a function of the system Fermi level ($E_{\rm F}$).}
We show results for NV defects near the core of 30$^{\circ}$ (90$^{\circ}$) dislocation in (a) [(b)]. Results for the NV in bulk diamond (NV@Bulk) are shown for comparison.
The slope of the curve indicates the stable charge state for a defect at a given $E_{\rm F}$ value, and the kink represents the charge transition level between two charge states for a defect.
The charge transition level of the N dopant in diamond, from the +1 to 0 charge state, i.e., N$_{\rm C}(+1/0)$, is indicated by the dashed vertical line.
See Fig.~1 for the NV notations.}
\label{s-fig:Ef_CTL}
\end{figure}

Based on the PBE-optimized geometries, we further performed single-point total energy calculations at the DDH level of theory in different charge states and for the corresponding ground-state spin state. We carried out equivalent calculations for the bulk NV in a 511-atom cubic supercell, to reduce the computational cost.
These total energies were used to obtain the formation energy $E_{\rm f}$ as a function of the system Fermi level ($E_{\rm F}$). The formation energy of an NV defect in charge state $q$ near the dislocation core, $E_{\rm f} ( {\rm NV}^q@{\rm disloc} )$, was computed as:
\begin{equation}
E_{\rm f} ( {\rm NV}^q@{\rm disloc} ) =
E_{\rm tot}( {\rm NV}^q@{\rm disloc} ) - E_{\rm tot}( {\rm disloc} )
- n_{\rm C}\mu_{\rm C} - n_{\rm N}\mu_{\rm N}
+ q E_{\rm F} + E_{\rm corr}
\end{equation}
where $E_{\rm tot}( {\rm NV}^q@{\rm disloc} )$ is the total energy of the supercell containing two dislocations with a relaxed ${\rm NV}^q$; $E_{\rm tot}( {\rm disloc} )$ is the total energy of the supercell with two dislocations without NV defect; $\mu_{\rm C}$ ($\mu_{\rm N}$) is the chemical potential of C (N) element; $n_{\rm C}$ ($n_{\rm N}$) the number of added C (N) atoms to form NV, which is $-2$ (+1); $E_{\rm F}$ is the Fermi energy referred to the valence band maximum; $E_{\rm corr}$ is the energy correction for spurious electrostatic interactions present in supercell calculations. We obtained $E_{\rm corr}$ using the method developed by Freysoldt, Neugebauer, and Van de Walle~\cite{freysoldt_fully_2009}; we used a dielectric constant of 5.7. The chemical potential $\mu_{\rm C}$ ($\mu_{\rm N}$) was calculated as the energy per atom in diamond (a nitrogen molecule). The total energies, $E_{\rm tot}( {\rm NV}^q@{\rm disloc} )$ and $E_{\rm tot}( {\rm disloc} )$, were obtained at the DDH level of theory based on relaxed geometries at the PBE level of theory as mentioned above. The calculation of the $E_{\rm f}$ for NV in bulk diamond in the 511-atom supercell is similar and not reported in detail here. These formation energy results for representative NVs near the dislocation core and NV in bulk diamond are presented in Fig.~\ref{s-fig:Ef_CTL}. Our computed formation energies and charge transition levels for NV in bulk diamond show good agreement with previous calculations using the HSE06 hybrid functional and projector-augmented wave (PAW) pseudopotentials~\cite{deak_formation_2014}.

We note some inconsistencies between our results and a previous computational study of NV centers in proximity to dislocations in diamond~\cite{ghassemizadeh_stability_2022}. For the G4-G9 defect near the 90$^{\circ}$ dislocation core, we found that the ground-state energy of its triplet spin state is lower by 0.18 eV than that of the singlet spin state, while this energy difference was found to be nearly zero (labeled as 2$_{i}$ in Ref.~\cite{ghassemizadeh_stability_2022}). In proximity to the 30$^{\circ}$ dislocation core, G9-G8 and G8-G9 exhibit slightly different energetics ($\Delta E_{\rm f}$), both being less stable than the G$^-$9-G11. However, G9-G8 and G8-G9 were reported to be identical and the most stable configurations (labeled as 1 and 1$_{i}$ respectively in Ref.~\cite{ghassemizadeh_stability_2022}).
In addition, we noticed significant structural reconstructions for some configurations near the dislocation core, which were not reported. One likely reason to explain these differences is that here we used a larger supercell and explored many more NV configurations as compared to Ref.~\cite{ghassemizadeh_stability_2022}. These observations suggest that, as we did in this work, a high-throughput-like method is helpful to accurately and fully characterize the behavior of spin defects near the dislocation core or other geometrically complex regions in the solid.

\section{Ground-state electronic structures of representative NV configurations}

\begin{figure}[!ht]
\centering
\includegraphics[width=0.7\textwidth]{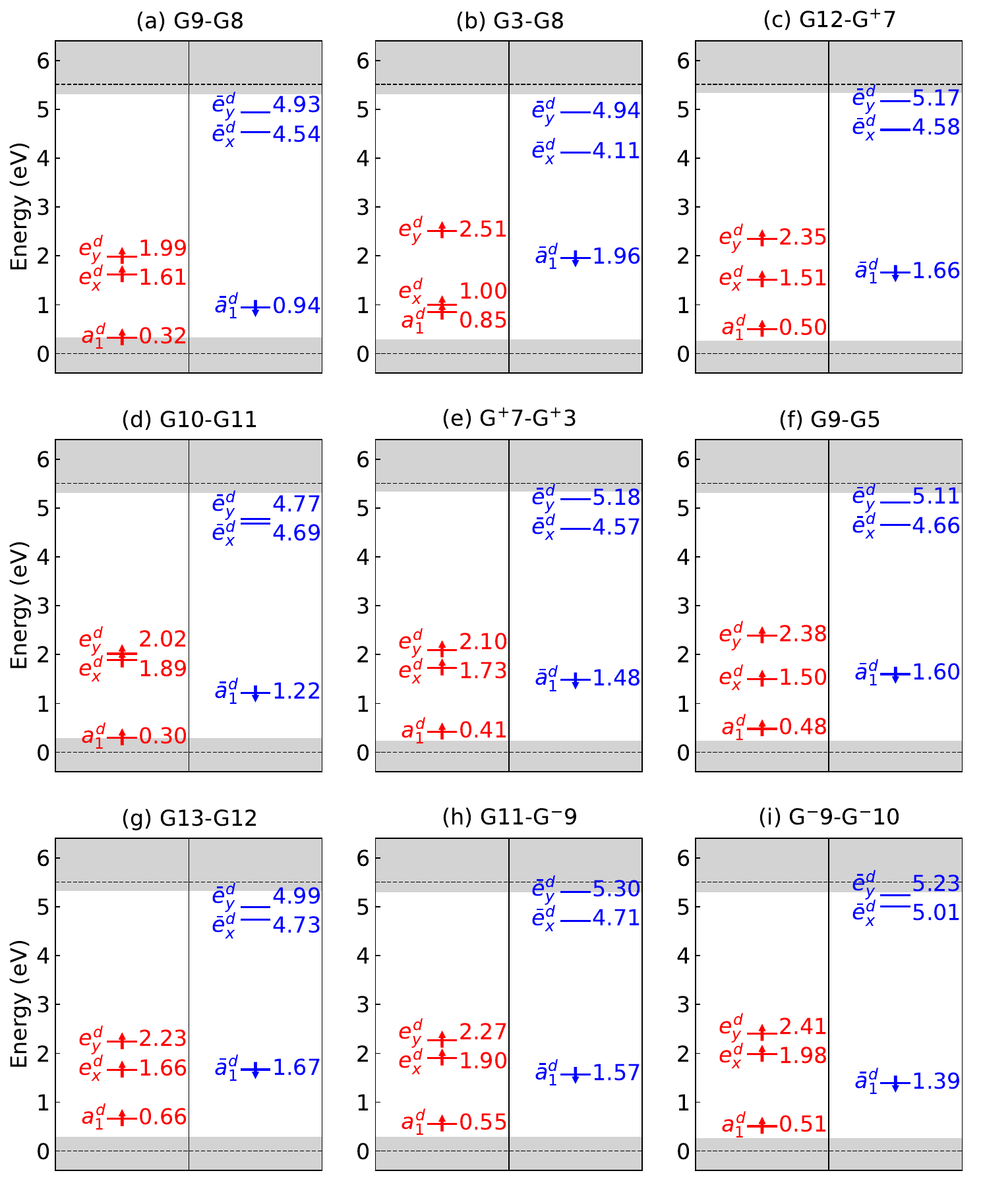}
\caption{
\textbf{Ground-state electronic structure of the representative NV defects near the 30$^{\circ}$ dislocation core.}
These results were computed for NV defects in the $-1$ charge state and the triplet spin state at the DDH level of theory based on geometries relaxed at the PBE level of theory.
The gray areas represent the valence and conduction bands of the 1727-atom diamond supercell with the 30$^{\circ}$ dislocations; the band edges of bulk diamond without dislocations are indicated by the horizontal dashed lines.
The single-particle defect levels in the up (down) spin channel are shown in red (blue); the colored number next to a defect level indicates its energy (in eV) referenced to the valence band maximum of bulk diamond.
See Fig.~1 for the NV notations.}
\label{s-fig:30DP_ES}
\end{figure}

\begin{figure}[!ht]
\centering
\includegraphics[width=0.7\textwidth]{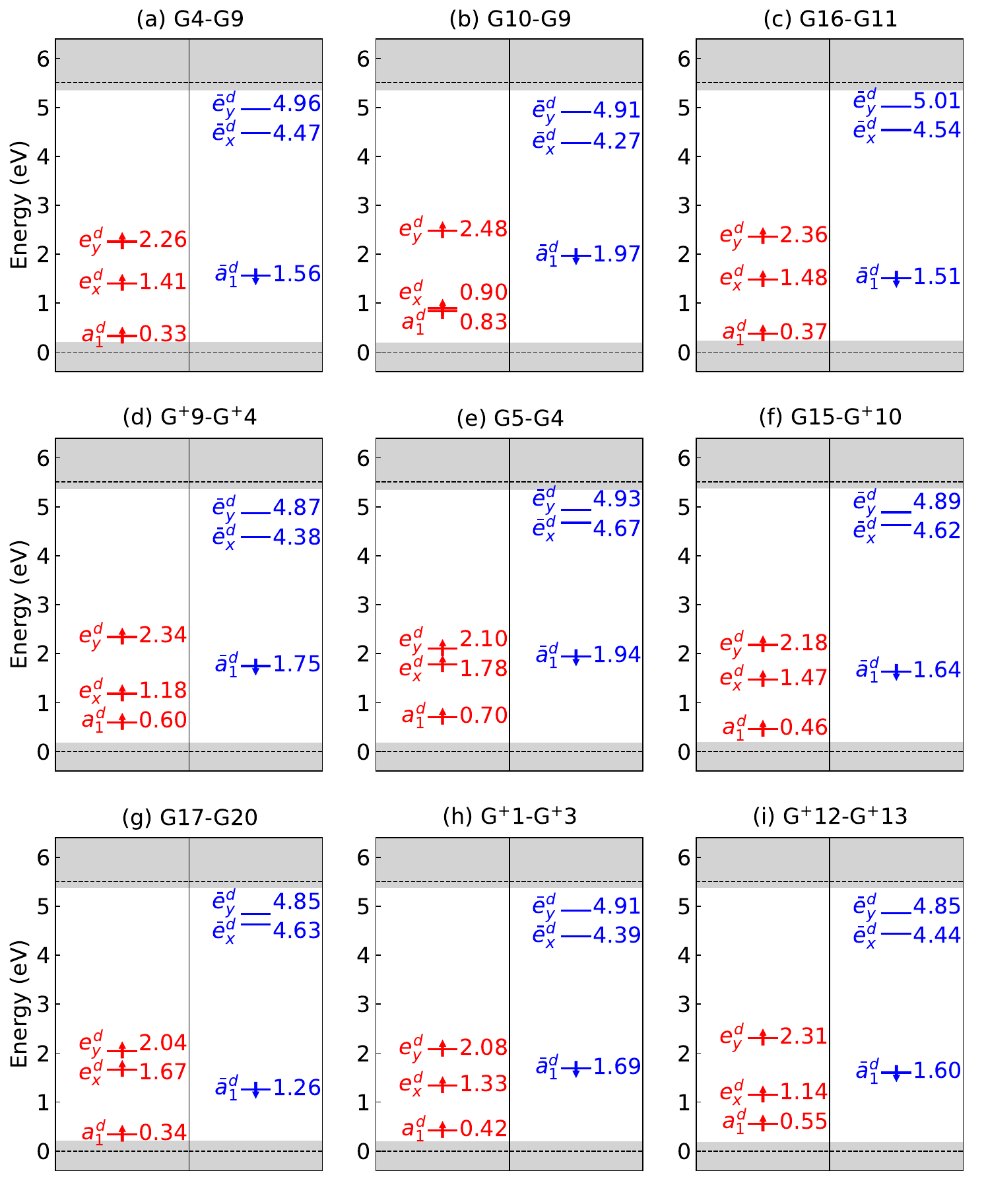}
\caption{
\textbf{Ground-state electronic structure of the representative NV defects near the 90$^{\circ}$ dislocation core.}
These results were computed for NV defects in the $-1$ charge state and the triplet spin state at the DDH level of theory based on geometries relaxed at the PBE level of theory.
The gray areas represent the valence and conduction bands of the 1727-atom diamond supercell with the 90$^{\circ}$ dislocations; the band edges of bulk diamond without dislocations are indicated by the horizontal dashed lines.
The single-particle defect levels in the up (down) spin channel are shown in red (blue); the colored number next to a defect level indicates its energy (in eV) referenced to the valence band maximum of bulk diamond.
See Fig.~1 for the NV notations.}
\label{s-fig:90DP_ES}
\end{figure}

We systematically investigated the electronic properties of the representative NVs near the core region. On top of the ground-state geometries relaxed at the PBE level of theory in the $-1$ charge state and the triplet spin state, we computed their total energies and electronic structures at the DDH level of theory, and extracted the defect levels within the gap of diamond, as shown in Figs.~\ref{s-fig:30DP_ES} and \ref{s-fig:90DP_ES}.

We used the localization factor (LF) to quantify the localization of the mid-gap defect states of the NVs near the dislocation core. The LF is defined as
\begin{equation}
{\rm LF}_i = \int_{\Omega_s} {\rm d}^3 \boldsymbol{r} | \psi_i (\boldsymbol{r}) | ^2
\end{equation}
where $\psi_i$ is the $i$th single-particle state obtained at the DDH level of theory. The integration was carried out within a sphere $\Omega_s$ centered around the carbon vacancy of the NV defect with a radius of 5 Bohr. The LF values of the representative NV configurations are shown in Table~\ref{s-tab:local_factor}. A large value of ${\rm LF}_i$ indicates that $\psi_i$ is localized around the vacancy and therefore is not hybridized with the dislocation-induced states or the bulk states of diamond.

\begin{table}[!ht]
\centering
\caption{
\textbf{Localization factor (LF) of ground-state single-particle defect states of the representative NV configurations near the 30$^{\circ}$ or 90$^{\circ}$ dislocation core.}
These results were computed at the DDH level of theory. A large value of LF indicates that the defect state is localized around the vacancy and therefore is not hybridized with the dislocation-induced states or bulk states of diamond.
See Fig.~1 for the NV notations.}
\footnotesize
\begin{tabular}{lccccccc}
\hline
\multirow{2}{*}{Dislocation} & \multirow{2}{*}{NV defect} & \multicolumn{6}{c}{LF} \\
\cline{3-8}
& & $a_1^d$ & $e_x^d$ & $e_y^d$ & $\bar{a}_1^d$ & $\bar{e}_x^d$ & $\bar{e}_y^d$ \\
\hline
None & NV@Bulk & 0.58 & 0.67 & 0.67 & 0.68 & 0.65 & 0.65 \\
\hline
\multirow{9}{*}{30$^{\circ}$}
& G9-G8 & 0.24 & 0.65 & 0.68 & 0.66 & 0.63 & 0.64 \\
& G3-G8 & 0.57 & 0.61 & 0.68 & 0.67 & 0.55 & 0.61 \\
& G12-G$^+$7 & 0.54 & 0.63 & 0.67 & 0.67 & 0.64 & 0.58 \\
& G10-G11 & 0.10 & 0.68 & 0.67 & 0.67 & 0.63 & 0.63 \\
& G$^+$7-G$^+$3 & 0.48 & 0.65 & 0.69 & 0.67 & 0.63 & 0.61 \\
& G9-G5 & 0.56 & 0.66 & 0.68 & 0.68 & 0.65 & 0.59 \\
& G13-G12 & 0.55 & 0.70 & 0.69 & 0.69 & 0.65 & 0.59 \\
& G11-G$^-$9 & 0.47 & 0.66 & 0.69 & 0.67 & 0.64 & 0.08 \\
& G$^-$9-G$^-$10 & 0.52 & 0.69 & 0.69 & 0.67 & 0.64 & 0.61 \\
\hline
\multirow{9}{*}{90$^{\circ}$}
& G4-G9 & 0.54 & 0.66 & 0.68 & 0.69 & 0.61 & 0.61 \\
& G10-G9 & 0.60 & 0.64 & 0.68 & 0.67 & 0.66 & 0.61 \\
& G16-G11 & 0.48 & 0.66 & 0.68 & 0.67 & 0.60 & 0.60 \\
& G$^+$9-G$^+$4 & 0.58 & 0.61 & 0.66 & 0.67 & 0.63 & 0.60 \\
& G5-G4 & 0.55 & 0.66 & 0.70 & 0.67 & 0.56 & 0.66 \\
& G15-G$^+$10 & 0.51 & 0.64 & 0.65 & 0.67 & 0.64 & 0.61 \\
& G17-G20 & 0.36 & 0.66 & 0.67 & 0.64 & 0.65 & 0.62 \\
& G$^+$1-G$^+$3 & 0.53 & 0.64 & 0.68 & 0.69 & 0.63 & 0.60 \\
& G$^+$12-G$^+$13 & 0.54 & 0.63 & 0.66 & 0.67 & 0.63 & 0.60 \\
\hline
\end{tabular}
\label{s-tab:local_factor}
\end{table}

\section{Vertical excitation energies of representative NV configurations}

\begin{figure}[!ht]
\centering
\includegraphics[width=0.7\textwidth]{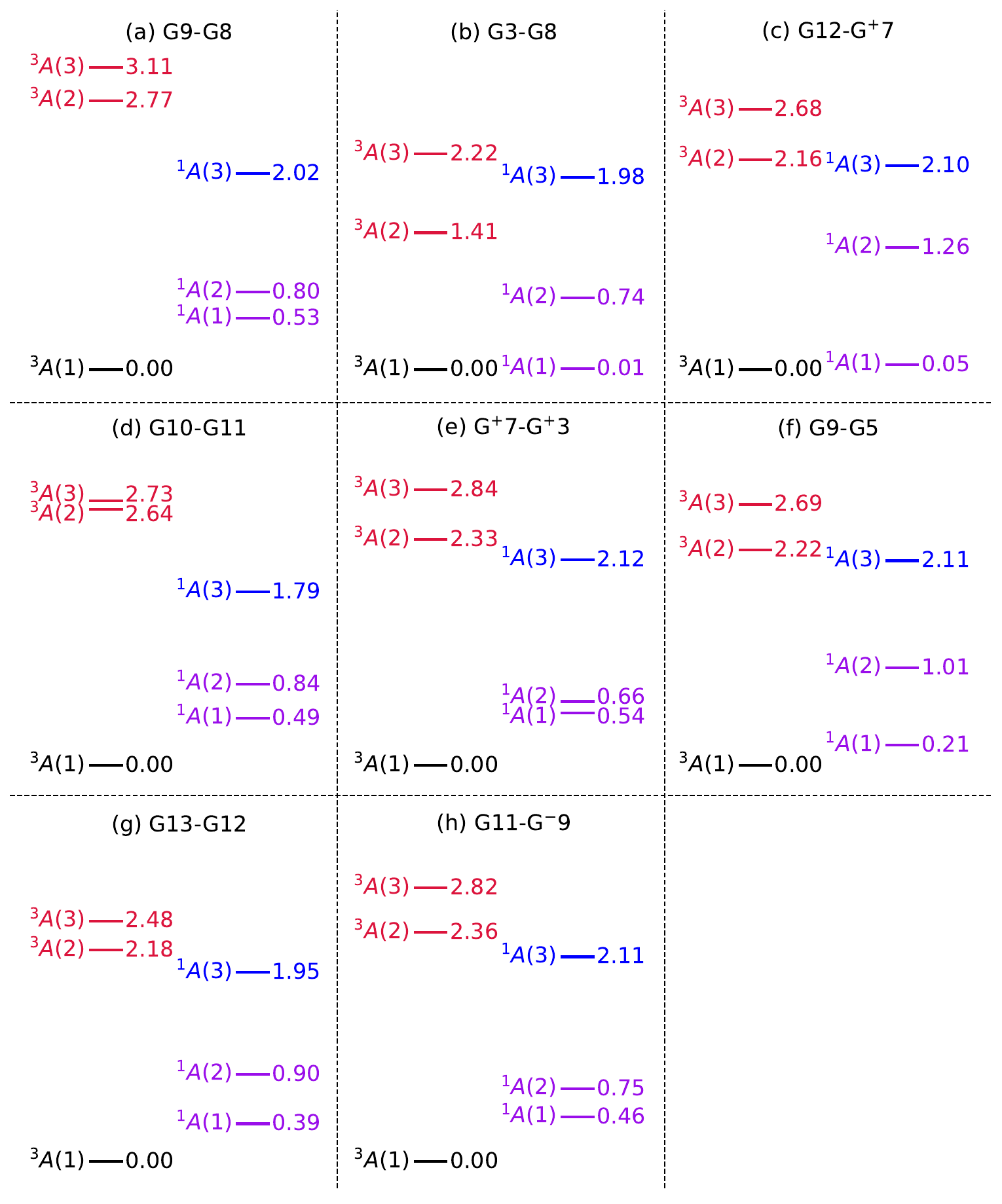}
\caption{
\textbf{Many-body electronic states of the representative NV defects in the $-1$ charge state near the 30$^{\circ}$ dislocation core.}
The numbers shown are the vertical excitation energies (VEEs, in eV) obtained using TDDFT calculations at the DDH level of theory based on geometries relaxed in the triplet ground state and the $-1$ charge state at the PBE level of theory.
The many-body triplet (singlet) states are labeled as $^3A$(1), $^3A$(2) and $^3A$(3) [$^1A$(1), $^1A$(2) and $^1A$(3)] in the order of increasing energy.
The result for the G$^-$9-G$^-$10 NV configuration is not shown, as its lowest spin-conserving excitations are defect-to-bulk transitions in the spin-up channel.
See Fig.~1 for the NV notations.}
\label{s-fig:30DP_VEE}
\end{figure}

\begin{figure}[!ht]
\centering
\includegraphics[width=0.7\textwidth]{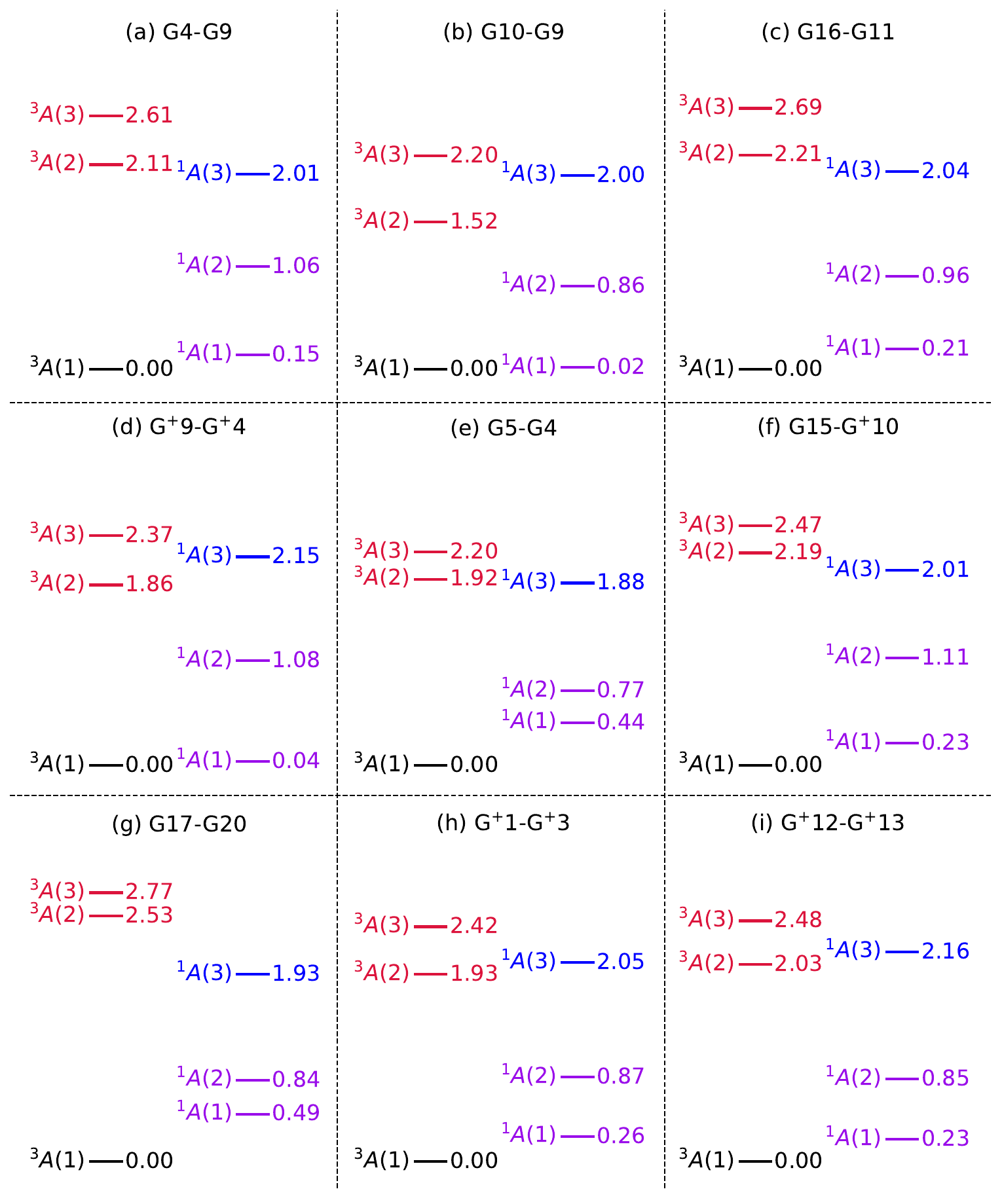}
\caption{
\textbf{Many-body electronic states of the representative NV defects in the $-1$ charge state near the 90$^{\circ}$ dislocation core.}
The numbers shown are the vertical excitation energies (VEEs, in eV) obtained using TDDFT calculations at the DDH level of theory based on geometries relaxed in the triplet ground state and the $-1$ charge state at the PBE level of theory.
The many-body triplet (singlet) states are labeled as $^3A$(1), $^3A$(2) and $^3A$(3) [$^1A$(1), $^1A$(2) and $^1A$(3)] in the order of increasing energy.
See Fig.~1 for the NV notations.}
\label{s-fig:90DP_VEE}
\end{figure}

To compute the vertical excitation energies of the lowest excited states, we used the WEST code~\cite{jin_excited_2023} to conduct linear-response time-dependent density functional theory (TDDFT) calculations within the Tamm-Dancoff approximation and with the DDH functional. The calculations were based on the ground-state geometries in the triplet spin state and the $-1$ charge state, relaxed using DFT at the PBE level of theory. Transitions from the triplet ground state to the triplet (spin-conserving) and singlet (spin-flip) excited states were considered. The results are presented in Figs.~\ref{s-fig:30DP_VEE} and \ref{s-fig:90DP_VEE}. We note that the use of DDH is critical as, for some NV configurations near the dislocation core, TDDFT calculations using the PBE functional incorrectly predicted the lowest excited state in the spin-up channel. In contrast, TDDFT calculations using the DDH functional yield the lowest excited state in the spin-down channel.

In Table~\ref{s-tab:VEE}, we compare the VEEs of the lowest defect-to-defect and defect-to-bulk spin-conserving excitations. For all representative NV configurations except the G$^-$9-G$^-$10 near the 30$^{\circ}$ dislocation, the intra-defect transition exhibits a lower VEE than the defect-to-bulk transition.

\begin{table}[!ht]
\centering
\caption{
\textbf{Vertical excitation energies (VEEs) of the representative NV defects near the dislocation core and NV in bulk diamond (NV@Bulk).}
These data were computed for NV in the $-1$ charge state using TDDFT calculations at the DDH level of theory based on the geometries relaxed at the PBE level of theory. We present the VEEs of the lowest defect-to-defect (D2D) and defect-to-bulk (D2B) spin-conserving vertical excitations.
See Fig.~1 for the NV notations.}
\footnotesize
\begin{tabular}{lccc}
\hline
Dislocation & NV defect & VEE D2D (eV) & VEE D2B (eV) \\
\hline
None & NV@Bulk & 2.28 & 3.37 \\
\hline
\multirow{9}{*}{30$^{\circ}$}
& G9-G8 & 2.77 & 3.12 \\
& G3-G8 & 1.41 & 2.58 \\
& G12-G$^+$7 & 2.16 & 2.76 \\
& G10-G11 & 2.64 & 3.07 \\
& G$^+$7-G$^+$3 & 2.33 & 2.99 \\
& G9-G5 & 2.22 & 2.71 \\
& G13-G12 & 2.18 & 2.85 \\
& G11-G$^-$9 & 2.36 & 2.82 \\
& G$^-$9-G$^-$10$^a$ & & 2.68 \\
\hline
\multirow{9}{*}{90$^{\circ}$}
& G4-G9 & 2.11 & 2.91 \\
& G10-G9 & 1.52 & 2.65 \\
& G16-G11 & 2.21 & 2.80 \\
& G$^+$9-G$^+$4 & 1.86 & 2.80 \\
& G5-G4 & 1.92 & 3.03 \\
& G15-G$^+$10 & 2.19 & 2.97 \\
& G17-G20 & 2.53 & 3.11 \\
& G$^+$1-G$^+$3 & 1.93 & 3.07 \\
& G$^+$12-G$^+$13 & 2.03 & 2.85 \\
\hline
\end{tabular}
\label{s-tab:VEE}
\begin{threeparttable}
\begin{tablenotes}
\footnotesize
\item[a] For the G$^-$9-G$^-$10 near the core of 30$^{\circ}$ dislocation, the lowest spin-conserving excitation is a D2B transition with a VEE of 2.68 eV.
\end{tablenotes}
\end{threeparttable}
\end{table}

\section{Optical properties of representative NV configurations}

The excited-state geometries of NV defects, corresponding to the first triplet excited state $^3A$(2) in Figs.~\ref{s-fig:30DP_VEE} and \ref{s-fig:90DP_VEE}, were optimized using constrained-occupation DFT (also called $\Delta$SCF) at the PBE level of theory. Excited-state geometries optimized using $\Delta$SCF and TDDFT, both with the PBE functional, are compared for two NV configurations in Table~\ref{s-tab:ES_relax}. The excited-state geometries relaxed with $\Delta$SCF are found to be close to those relaxed with TDDFT.

\begin{table}[!ht]
\centering
\caption{
\textbf{Comparison of excited-state [$^3A$(2) in Figs.~\ref{s-fig:30DP_VEE} and \ref{s-fig:90DP_VEE}] geometries optimized using $\Delta$SCF and TDDFT, both with the PBE functional.}
For the excited-state geometries, we present the total energies ($E_{\rm tot}$) and the mass-weighted displacements ($\Delta Q$, defined in Eq.~\ref{s-eq:delta_q}).
See Fig.~1 for the NV notations.}
\footnotesize
\begin{tabular}{lccc}
\hline
NV defect & Method & $E_{\rm tot}$ (eV) & $\Delta Q$ (amu$^{1/2}$ \AA) \\
\hline
\multirow{2}{*}{G3-G8@30$^{\circ}$} & $\Delta$SCF & -267696.407 & 0.706 \\
& TDDFT & -267696.402 & 0.728 \\
\hline
\multirow{2}{*}{G10-G9@90$^{\circ}$} & $\Delta$SCF & -267688.775 & 0.751 \\
& TDDFT & -267688.746 & 0.721 \\
\hline
\end{tabular}
\label{s-tab:ES_relax}
\end{table}

For the representative NVs near the core region, we computed their zero-phonon line (ZPL), Huang-Rhys factor (HRF), and Debye-Waller factor (DWF) for the transition from the first excited triplet to the ground triplet state. The DWF describes the ratio between the emission intensity of the ZPL and that of the entire photoluminescence (PL) spectrum, reflecting the strength of electron-phonon coupling for the optical transition. The ZPL energy for an NV defect was computed as:
\begin{equation}
E_{\rm ZPL} = E_{\rm tot}^{\rm ES} - E_{\rm tot}^{\rm GS} + E_{\rm V}^{\rm ES}
\end{equation}
where $E_{\rm tot}^{\rm GS}$ ($E_{\rm tot}^{\rm ES}$) is the ground-state total energy of the supercell containing an NV defect with atomic coordinates relaxed in the ground (first excited) triplet state; $E_{\rm V}^{\rm ES}$ is the VEE for the ground to first triplet excited state transition computed with TDDFT at the NV configuration relaxed in the first excited triplet state. $E_{\rm tot}^{\rm ES}$, $E_{\rm tot}^{\rm GS}$ and $E_{\rm V}^{\rm ES}$ were computed at the DDH level of theory. The relaxed ground (first excited) triplet state NV configurations were obtained by the DFT ($\Delta$SCF) calculations at the PBE level of theory. Our computed ZPL energy of 2.07 eV for the NV in bulk diamond is close to the experimentally measured value of 1.945 eV.

This ZPL transition can be approximately described by a one-dimensional (1D) phonon model. Specifically, for a given NV defect, based on the relaxed atomic coordinates of the ground ($\bm{{\rm R}}_{\rm GS}$) and first excited ($\bm{{\rm R}}_{\rm ES}$) triplet state, the scaled coordinate difference was computed as $\Delta \bm{{\rm R}} = (\bm{{\rm R}}_{\rm ES} - \bm{{\rm R}}_{\rm GS}) / \Delta Q$, with $\Delta Q$ being the mass-weighted displacement calculated as:
\begin{equation}
\Delta Q = \sqrt{ \sum_{i=1}^{N} m_{i} \times \sum_{\alpha}(\bm{{\rm R}}_{\rm ES}^{i\alpha} - \bm{{\rm R}}_{\rm GS}^{i\alpha})^{2} }
\label{s-eq:delta_q}
\end{equation}
where $i$ is the atomic index; $N$ is the total number of atoms; $\alpha=x,y,z$ directions; $m_{i}$ is the mass of the $i$th atom; $\bm{{\rm R}}_{\rm GS}^{i\alpha}$ ($\bm{{\rm R}}_{\rm ES}^{i\alpha}$) is the atomic coordinate along the $\alpha$ direction of the $i$th atom in the relaxed ground (first excited) triplet state configuration.

\begin{figure}[!ht]
\centering
\includegraphics[width=0.7\textwidth]{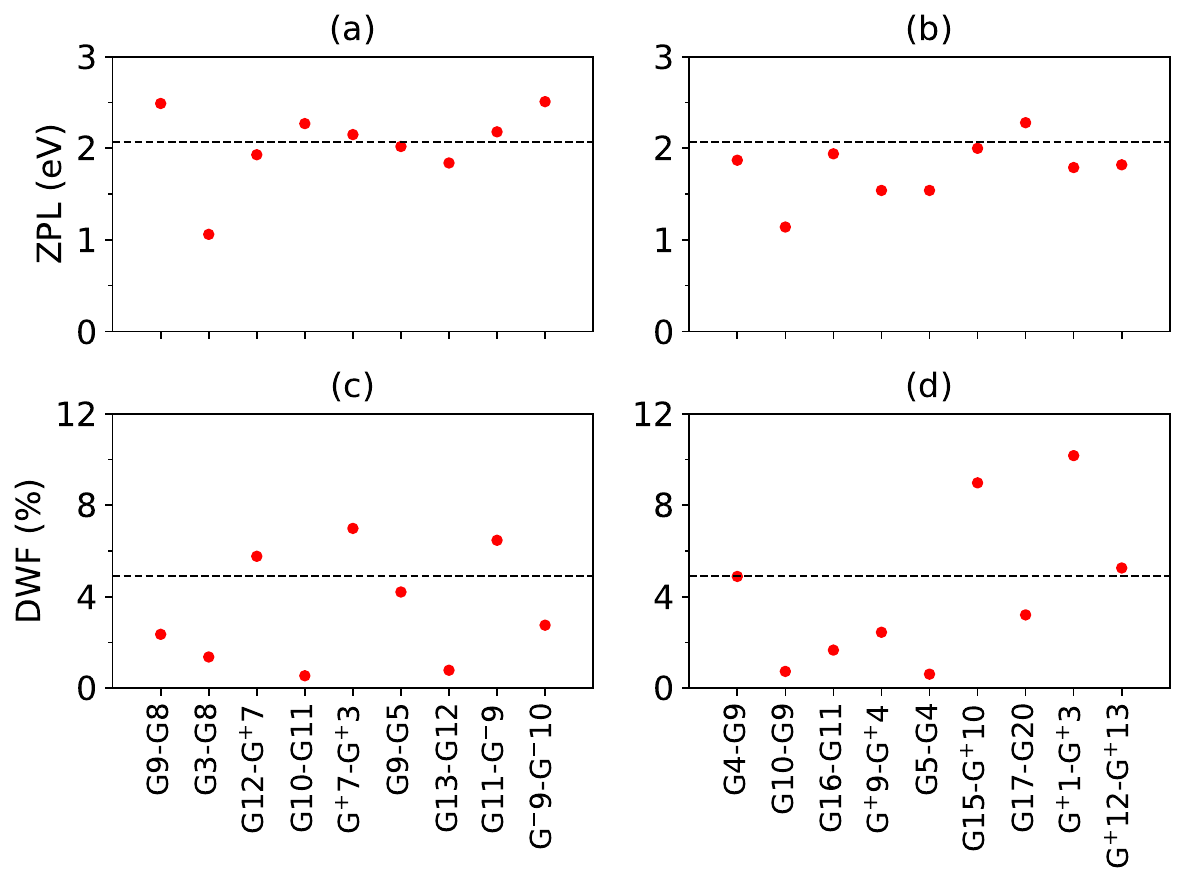}
\caption{
\textbf{Computed zero-phonon line (ZPL) energies and Debye-Waller factors (DWF) for the representative NV defects in the $-1$ charge state.}
The ZPL values for NV near the core of 30$^{\circ}$ (90$^{\circ}$) dislocation are shown in (a) [(b)]. The DWF values for NV near the core of 30$^{\circ}$ (90$^{\circ}$) dislocation are shown in (c) [(d)]. The calculated value for NV in bulk diamond is indicated by the dashed line.
See Fig.~1 for the NV notations.}
\label{s-fig:ZPL}
\end{figure}

We then linearly interpolated a series of configurations as $\bm{{\rm R}}_{\rm int}(Q) = \bm{{\rm R}}_{\rm GS} + \Delta \bm{{\rm R}} \times Q$, where $Q$ goes from $-0.2\Delta Q$ to 1.2$\Delta Q$. For each interpolated structure $\bm{{\rm R}}_{\rm int}(Q)$, we computed its total energy $E_{\rm GS}(Q)$ [$E_{\rm ES}(Q)$] in the ground (first excited) triplet state at the PBE level of theory using DFT ($\Delta$SCF). The effective phonon frequencies were obtained by fitting a quadratic function around the respective minima of $E_{\rm GS}(Q)$ and $E_{\rm ES}(Q)$ curves, denoted as $\Omega_{\rm GS}$ and $\Omega_{\rm ES}$. Then, we computed the HRF $= \Omega_{\rm GS}\Delta Q^{2}/(2\hbar)$, and the DWF $= \exp(-{\rm HRF})$; the results obtained using $\Omega_{\rm ES}$ are slightly different, and not reported here as the differences are negligible. We note that our computed HRF value of 3.02 for the NV in bulk diamond agrees with the value of 3.22 reported in a previous study~\cite{jin_photoluminescence_2021}. The computed ZPL and DFW values are presented in Fig.~\ref{s-fig:ZPL}.

\section{Inter-system crossing rates}

Inter-system crossing (ISC) rates $\Gamma$ were calculated using a recently developed computational framework that has been demonstrated to accurately predict ISC rates for NVs in bulk diamond at temperatures below 600 K~\cite{jin_first_2025}. Using the Fermi's golden rule,
\begin{equation}
\Gamma = \frac{2\pi}{\hbar} |\lambda|^2 F(\Delta).
\end{equation}
For the upper ISC process in the NV center in bulk diamond, namely the $^3\!E \to ^1\!A_1$ transition, $\lambda = \braket{^3\!E_{M_S = \pm 1} | \hat{H}_{\mathrm{SOC}} | ^1\!A_1}$ denotes the spin-orbit coupling (SOC) matrix element between the $M_S = \pm 1$ sub-levels of the triplet $^3\!E$ state and the singlet $^1\!A_1$ state. The term $F(\Delta)$ is the vibrational overlap function for this ISC pathway, where $\Delta$ represents the zero-phonon energy gap between the $^3\!E$ and $^1\!A_1$ states. For the lower ISC process, corresponding to the $^1\!E \to ^3\!A_2$ transition, $\lambda = \braket{^1\!E | \hat{H}{\mathrm{SOC}} | ^3\!A_{2, M_S = \pm 1}}$ is the SOC matrix element between the $^1\!E$ state and the $M_S = \pm 1$ sub-levels of the $^3\!A_2$ ground state. The corresponding vibrational overlap function for this transition is denoted as $H(\Sigma)$, where $\Sigma$ is the zero-phonon gap between $^1\!E$ and $^3\!A_2$. According to recent theoretical predictions, $\Delta$ and $\Sigma$ of the NV in bulk diamond are approximately 0.355 eV and 0.400 eV, respectively~\cite{jin_first_2025}.

In this work, rather than directly computing the ISC rates of NV defects located near dislocation cores, we estimate them using the experimentally determined ISC rates of NV defects in bulk diamond as a reference~\cite{robledo_spin_2011}. Specifically, we evaluate the changes in the SOC matrix elements ($\lambda$), vibrational overlap functions [$F(\Delta)$ and $H(\Sigma)$], and zero-phonon energy gaps ($\Delta$ and $\Sigma$) for NV defects near dislocations relative to their bulk counterparts. The upper ISC rate for an NV center near a dislocation is then given by:
\begin{equation}
\Gamma_{\mathrm{NV}@\mathrm{Dislocation}} = \Gamma_{\mathrm{NV}@\mathrm{Bulk}} \left| \dfrac{\lambda_{\mathrm{NV}@\mathrm{Dislocation}}}{\lambda_{\mathrm{NV}@\mathrm{Bulk}}} \right|^2 \dfrac{F_{{\mathrm{NV}@\mathrm{Dislocation}}}(\Delta_{{\mathrm{NV}@\mathrm{Dislocation}}})}{F_{{\mathrm{NV}@\mathrm{Bulk}}}(\Delta_{{\mathrm{NV}@\mathrm{Bulk}}})} ,
\end{equation}
and similarly for the lower ISC rate. The SOC matrix elements ($\lambda_{\mathrm{NV}@\mathrm{Dislocation}}$ and $\lambda_{\mathrm{NV}@\mathrm{Bulk}}$) were computed using many-body wavefunctions obtained from quantum defect embedding theory (QDET) calculations~\cite{sheng_green_2022,chen_advances_2025}. The one-body SOC operators are constructed as the difference between fully relativistic and scalar relativistic norm-conserving pseudopotentials~\cite{jin_first_2025}. For NV defects located near dislocation cores, where symmetry breaking leads to non-zero transverse $E$ terms in the ground-state zero-field splitting (ZFS), we apply a basis transformation from the $\ket{\pm 1}$ spin states to the $\ket{\pm}$ basis, defined as $\ket{+} = \left(\ket{+1} + \ket{-1}\right)/\sqrt{2}$ and $\ket{-} = \left(\ket{+1} - \ket{-1}\right)/\sqrt{2}$.

\begin{figure}[!ht]
\centering
\includegraphics[width=0.7\textwidth]{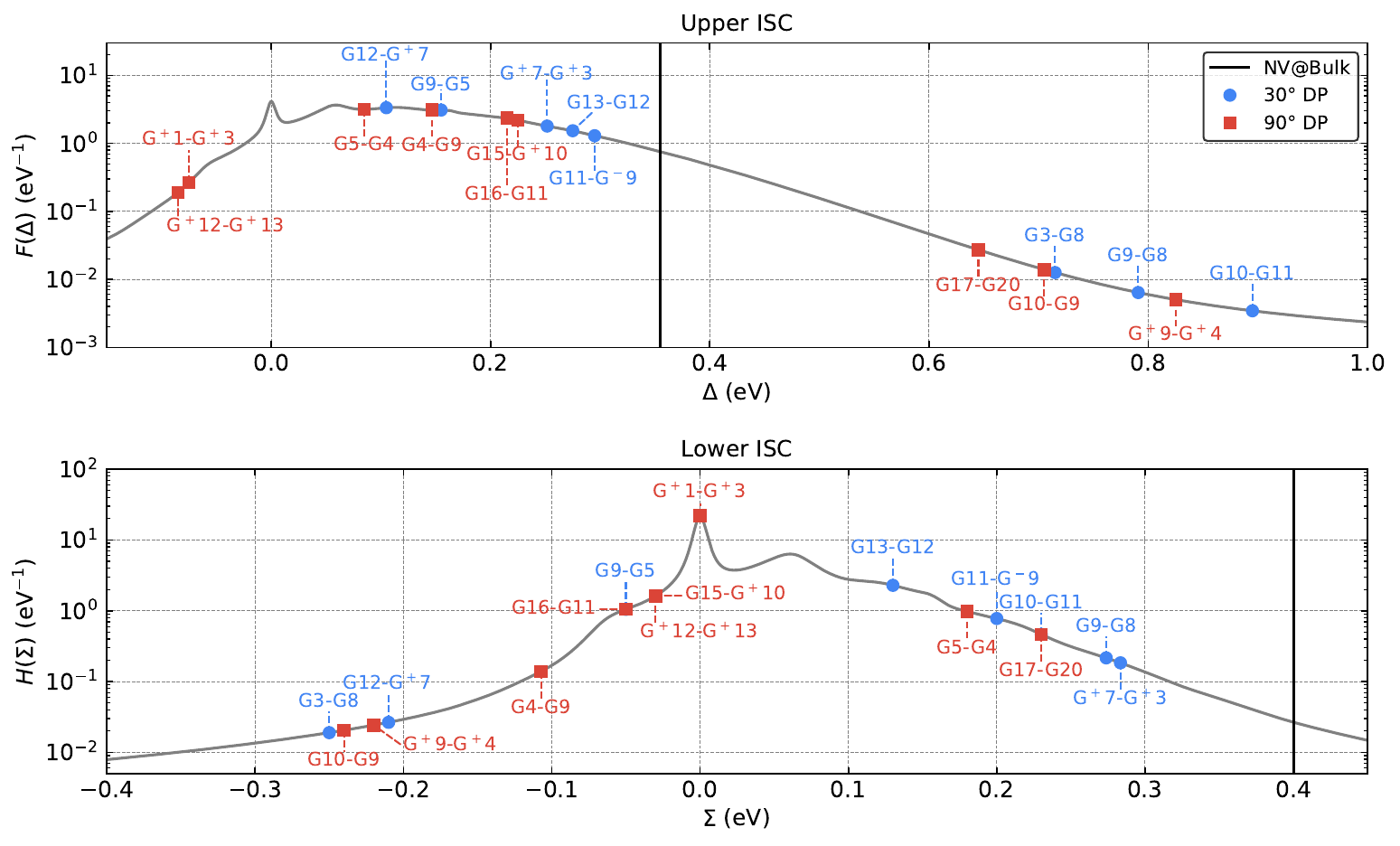}
\caption{
\textbf{Vibrational overlap functions (VOFs) for the upper inter-system crossing (ISC) transition [$F(\Delta)$] and the lower ISC transition [$H(\Sigma)$].}
The VOFs are plotted as a function of the zero-phonon energy gap between the triplet excited state and the higher-energy singlet state ($\Delta$), or the gap between the lower-energy singlet state and the triplet ground state ($\Sigma$). The theoretically estimated values of $\Delta$ and $\Sigma$ for the NV center in bulk diamond are indicated by vertical solid lines~\cite{jin_first_2025}. The corresponding values for NV defects near dislocation cores are shown as blue circles (30$^{\circ}$) and red squares (90$^{\circ}$).
See Fig.~1 for the NV notations.}
\label{s-fig:spectral_fxns}
\end{figure}

The vibrational overlap functions $F(\Delta)$ and $H(\Sigma)$ for the NV center in bulk diamond were computed using the Huang-Rhys theory and the generating function approach~\cite{jin_photoluminescence_2021}. The equilibrium geometries of the triplet and singlet excited states are obtained from spin-conserving and spin-flip TDDFT calculations, respectively~\cite{jin_excited_2023}. To account for the slight inaccuracy in the geometry relaxation associated with the use of the PBE functional in describing the $^3\!E \to ^1\!A_1$ (upper ISC) and $^1\!E \to ^3\!A_2$ (lower ISC) transitions, we apply scaling factors to the partial HRFs as in our previous work ~\cite{jin_vibrationally_2022}. The resulting vibrational overlap functions, evaluated at 300 K, are shown in Fig.~\ref{s-fig:spectral_fxns}.

For NV centers near dislocation cores, direct calculation of vibrational overlap functions is computationally prohibitive due to the cost of relaxing excited-state geometries and evaluating phonon modes for singlet states in the 1727-atom supercell. Therefore, we assume the vibrational overlap functions for these configurations are approximately equal, in order of magnitude, to those of the NV center in bulk diamond. This approximation is supported by the comparable mass-weighted displacements ($\Delta Q$) computed for the triplet excited states in both defect environments.

The zero-phonon energy gaps $\Delta$ and $\Sigma$ were computed using TDDFT with the DDH functional, and the results are shown in Figs.~\ref{s-fig:30DP_VEE} and \ref{s-fig:90DP_VEE}. To account for geometry relaxation effects and to align the computed values with theoretical estimates, 0.355 eV for $\Delta$ and 0.400 eV for $\Sigma$~\cite{jin_first_2025}, rigid energy shifts are applied. Specifically, a shift of +0.045 eV is added to $\Delta$, and a shift of $-0.26$ eV is applied to $\Sigma$.

In this study, to simplify the treatment of ISC transitions for NV defects near dislocation cores, we consider only the transitions from the lower-energy branch of the triplet excited state to the higher-energy singlet excited state [$^3A$(2) $\rightarrow$ $^1A$(3)], and from the lower-energy branch of the singlet state to the triplet ground state [$^1$A(1) $\rightarrow$ $^3A$(1)]. We assume that non-radiative transitions within the triplet excited manifold and within the singlet manifold are much faster than the ISC processes, due to the small energy gaps between these states.

In addition to the ISC rates explicitly computed and presented in Fig.~3 of the main text, Fig.~\ref{s-fig:spectral_fxns} provides insights into the NVs that may exhibit desirable ISC rates, based on their zero-phonon energy gaps ($\Delta$ and $\Sigma$) and vibrational overlap functions. For the upper ISC process, configurations with a $\Delta$ 0.2 eV larger than that of the NV center in bulk diamond may exhibit ISC rates too slow to support efficient optical initialization and readout. One such example is the G9-G8@30$^{\circ}$ configuration, as discussed in the main text. Conversely, configurations with a negative $\Delta$ may still be optically active at room temperature due to thermal population of higher vibronic levels. For the lower ISC process, all configurations exhibit smaller $\Sigma$ values compared to the bulk NV center, which, as discussed in the main text, suggests potentially faster ISC rates. However, configurations with a negative $\Sigma$ may raise concerns regarding the relative stability of the triplet and singlet ground states. Considering both the upper and lower ISC processes, the configurations that are most likely to exhibit favorable ISC properties include G$^+$7-G$^+$3, G13-G12, and G11-G$^-$9 at the 30$^{\circ}$ dislocation, and G5-G4 at the 90$^{\circ}$ dislocation. This inference is supported by explicit ISC calculations for G$^+$7-G$^+$3 and G13-G12 at the 30$^{\circ}$ dislocation.

\section{State population simulations}

\begin{figure}[!ht]
\centering
\includegraphics[width=0.5\textwidth]{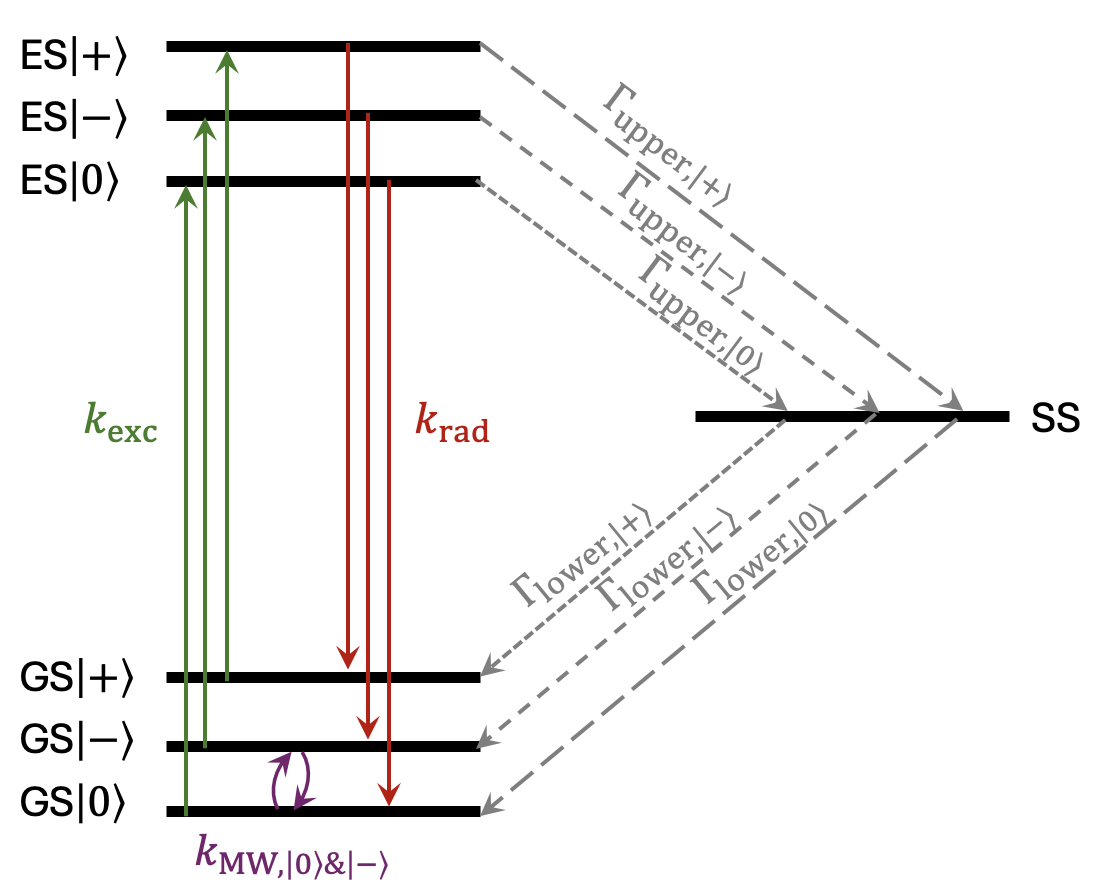}
\caption{
\textbf{Schematic diagram illustrating the energy levels and kinetic processes of the NV center in diamond.}
The excitation and radiative decay rates between the triplet states are denoted by $k_{\mathrm{exc}}$ and $k_{\mathrm{rad}}$, respectively. The inter-system crossing (ISC) rates from and to the $\ket{+}$ sub-level are represented by $\Gamma_{\mathrm{upper},\ket{+}}$ and $\Gamma_{\mathrm{lower},\ket{+}}$, respectively; analogous rates apply for the $\ket{-}$ and $\ket{0}$ sub-levels. The microwave-driven transition between the $\ket{0}$ and $\ket{-}$ sub-levels of the ground state is labeled as $k_{\mathrm{MW},\ket{0}\&\ket{-}}$; the corresponding transition between the $\ket{0}$ and $\ket{+}$ sub-levels is omitted for clarity.}
\label{s-fig:7_level_model}
\end{figure}

A seven-state model is employed in the state population simulations, comprising three spin sub-levels of the triplet ground state (GS): GS$\ket{+}$, GS$\ket{-}$, and GS$\ket{0}$ (corresponding to GS$\ket{+1}$, GS$\ket{-1}$, and GS$\ket{0}$ for the NV defect in bulk diamond), and three spin sub-levels of the triplet excited state (ES): ES$\ket{+}$, ES$\ket{-}$, and ES$\ket{0}$ (ES$\ket{+1}$, ES$\ket{-1}$, and ES$\ket{0}$ in bulk). In addition, a contracted singlet state (SS) is included. The full model is illustrated in Fig.~\ref{s-fig:7_level_model}.

The time evolution of the population in each of the seven states is governed by the following set of coupled rate equations:
\begin{equation}
\begin{aligned}
\dfrac{d P_{\mathrm{GS}\ket{0}}(t)}{dt} &= - \left(k_{\mathrm{exc}} - k_{\mathrm{MW}, \ket{0} \& \ket{-}} - k_{\mathrm{MW}, \ket{0}\&\ket{+}}\right) P_{\mathrm{GS}\ket{0}}(t) \\
& + k_{\mathrm{MW}, \ket{0}\&\ket{-}} P_{\mathrm{GS}\ket{-}}(t) + k_{\mathrm{MW}, \ket{0}\&\ket{+}} P_{\mathrm{GS}\ket{+}}(t) \\
& + k_{\mathrm{rad}} P_{\mathrm{ES}\ket{0}}(t) + \Gamma_{\mathrm{lower}, \ket{0}} P_{\mathrm{SS}}(t),
\end{aligned}
\end{equation}
\begin{equation}
\begin{aligned}
\dfrac{d P_{\mathrm{GS}\ket{-}}(t)}{dt} &= k_{\mathrm{MW}, \ket{0} \& \ket{-}} P_{\mathrm{GS}\ket{0}}(t) - \left(k_{\mathrm{exc}} - k_{\mathrm{MW}, \ket{0} \& \ket{-}} \right) P_{\mathrm{GS}\ket{-}}(t) \\
& + k_{\mathrm{rad}} P_{\mathrm{ES}\ket{-}}(t) + \Gamma_{\mathrm{lower}, \ket{-}} P_{\mathrm{SS}}(t),
\end{aligned}
\end{equation}
\begin{equation}
\begin{aligned}
\dfrac{d P_{\mathrm{GS}\ket{+}}(t)}{dt} &= k_{\mathrm{MW}, \ket{0} \& \ket{+}} P_{\mathrm{GS}\ket{0}}(t) - \left(k_{\mathrm{exc}} - k_{\mathrm{MW}, \ket{0} \& \ket{+}} \right) P_{\mathrm{GS}\ket{+}}(t) \\
& + k_{\mathrm{rad}} P_{\mathrm{ES}\ket{+}}(t) + \Gamma_{\mathrm{lower}, \ket{+}} P_{\mathrm{SS}}(t),
\end{aligned}
\end{equation}
\begin{equation}
\begin{aligned}
\dfrac{d P_{\mathrm{ES}\ket{0}}(t)}{dt} &= k_{\mathrm{exc}} P_{\mathrm{GS}\ket{0}}(t) - \left(k_{\mathrm{rad}} + \Gamma_{\mathrm{upper}, \ket{0}} \right) P_{\mathrm{ES}\ket{0}}(t),
\end{aligned}
\end{equation}
\begin{equation}
\begin{aligned}
\dfrac{d P_{\mathrm{ES}\ket{-}}(t)}{dt} &= k_{\mathrm{exc}} P_{\mathrm{GS}\ket{-}}(t) - \left(k_{\mathrm{rad}} + \Gamma_{\mathrm{upper}, \ket{-}} \right) P_{\mathrm{ES}\ket{-}}(t),
\end{aligned}
\end{equation}
\begin{equation}
\begin{aligned}
\dfrac{d P_{\mathrm{ES}\ket{+}}(t)}{dt} &= k_{\mathrm{exc}} P_{\mathrm{GS}\ket{+}}(t) - \left(k_{\mathrm{rad}} + \Gamma_{\mathrm{upper}, \ket{+}} \right) P_{\mathrm{ES}\ket{+}}(t),
\end{aligned}
\end{equation}
\begin{equation}
\begin{aligned}
\dfrac{d P_{\mathrm{SS}}(t)}{dt} &= \Gamma_{\mathrm{upper}, \ket{0}} P_{\mathrm{ES}\ket{0}}(t) + \Gamma_{\mathrm{upper}, \ket{-}} P_{\mathrm{ES}\ket{-}}(t) + \Gamma_{\mathrm{upper}, \ket{+}} P_{\mathrm{ES}\ket{+}}(t) \\
& - \left( \Gamma_{\mathrm{lower}, \ket{0}} + \Gamma_{\mathrm{lower}, \ket{-}} + \Gamma_{\mathrm{lower}, \ket{+}} \right) P_{\mathrm{SS}}(t).
\end{aligned}
\end{equation}
In the simulations, we use a radiative decay rate of $k_{\mathrm{rad}} = 62.5$ GHz~\cite{robledo_spin_2011}. The excitation rate is denoted as $k_{\mathrm{exc}}$, and the microwave-driven transition rates between ground-state sub-levels are denoted as $k_{\mathrm{MW},\ket{0} \& \ket{-}}$ and $k_{\mathrm{MW},\ket{0} \& \ket{+}}$. The time-dependent populations $P$ of all states were computed analytically by applying the matrix exponential of the rate matrix multiplied by the initial population vector.

For the optical initialization simulation, we assume an initial population equally distributed among the triplet ground-state sub-levels: $P_{\mathrm{GS}\ket{0}}(t=0) = P_{\mathrm{GS}\ket{-}}(t=0) = P_{\mathrm{GS}\ket{+}}(t=0) = 1/3$, and use an excitation rate of $k_{\mathrm{exc}} = 0.1$ GHz. The populations of all states were computed at various time points and shown in Fig.~4 of the main text.

For the continuous-wave optically detected magnetic resonance (cw-ODMR) contrast simulation, the same initial population is used for both the microwave (MW) on and off conditions. The excitation rate is set to $k_{\mathrm{exc}} = 0.1$ GHz. To simulate the MW resonance between GS$\ket{0}$ and GS$\ket{-}$, we set $k_{\mathrm{MW},\ket{0}\&\ket{-}} = 1$ GHz and $k_{\mathrm{MW},\ket{0}\&\ket{+}} = 0$ GHz; for the resonance between GS$\ket{0}$ and GS$\ket{+}$, we set $k_{\mathrm{MW},\ket{0}\&\ket{-}} = 0$ GHz and $k_{\mathrm{MW},\ket{0}\&\ket{+}} = 1$ GHz. The ODMR contrast is computed as
\begin{equation}
\frac{\Delta I_{\mathrm{PL}}}{I_{\mathrm{PL}}} = 1 - \frac{P_{\mathrm{ES,\ Total,\ w/\ MW}}(t_{\mathrm{steady}})}{P_{\mathrm{ES,\ Total,\ w/o\ MW}}(t_{\mathrm{steady}})},
\end{equation}
which quantifies the relative change in the total excited-state population at steady state in the presence or absence of MW driving. The results are presented in Fig.~4, where the ODMR peaks are represented as Lorentzian functions centered at $D-E$ and $D+E$ for the $|-\rangle$ and $|+\rangle$ spin sub-levels, respectively. Here $D$ and $E$ denote the computed axial and transverse components of the ZFS tensor (Fig.~\ref{s-fig:ZFS}). An artificial broadening of 0.05 GHz is applied, and the peak amplitudes are determined based on the results of the state population simulations.

For the PL spin-contrast readout simulations, we initialize the system in one of the triplet ground-state sub-levels: $P_{\mathrm{GS}\ket{0}}(t=0) = 1$, $P_{\mathrm{GS}\ket{-}}(t=0) = 1$, or $P_{\mathrm{GS}\ket{+}}(t=0) = 1$, and use an excitation rate of $k_{\mathrm{exc}} = 12.8$ GHz. The total excited-state population is computed for the first 10,000 ns, and the results are shown in Fig.~4. The choice of parameters in all simulations is intended to provide a representative behavior and to illustrate the relevant contrast mechanisms.

\section{Zero-field splitting and hyperfine interaction parameters of representative NV configurations}

\begin{figure}[!ht]
\centering
\includegraphics[width=0.7\textwidth]{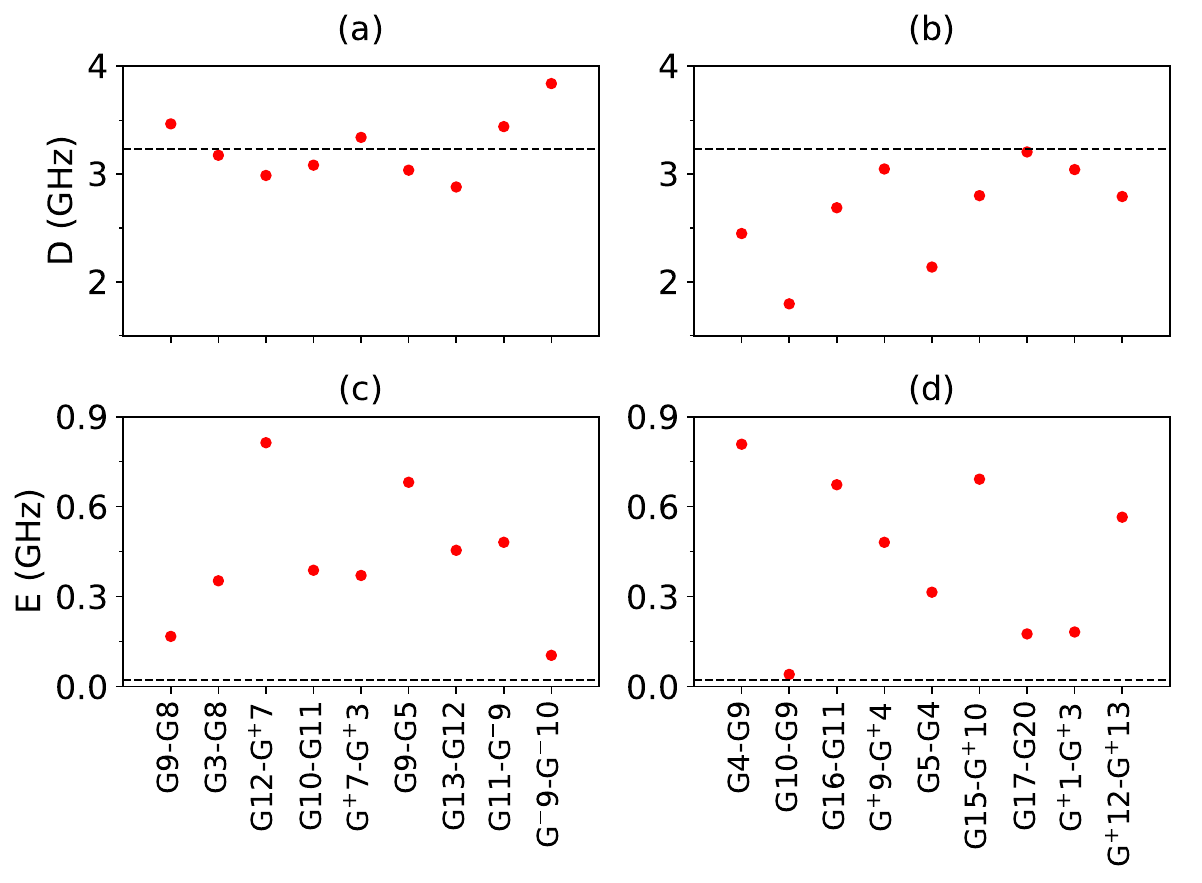}
\caption{
\textbf{Computed axial ($D$) and transverse ($E$) components of the zero-field splitting tensor for the representative NVs in the $-1$ charge state.}
The $D$ values for NV near the core of 30$^{\circ}$ and 90$^{\circ}$ dislocations are shown in (a) and (b), and the $E$ values in (c) and (d), respectively. The calculated value for NV in bulk diamond is indicated by the dashed line.
The $E$ value for the NV in bulk diamond should be zero due to the $C_{3v}$ symmetry, while a small value of 21.27 MHz was obtained numerically; this is a side effect of using the non-cubic 1728-atom supercell in our calculation (Fig.~\ref{s-fig:supercell}).
See Fig.~1 for the NV notations.}
\label{s-fig:ZFS}
\end{figure}

Using the supercell of diamond with dislocations and various orientations of the defect near the dislocation cores, we computed the electron-nuclear hyperfine interactions using DFT at the PBE level of theory as implemented in the GIPAW module of Quantum ESPRESSO~\cite{qe-gipaw}; pseudopotentials constructed for GIPAW were used. The results for representative NVs near the core region and the NV in bulk diamond are shown in Table~\ref{s-tab:hyperfine}. The interactions between the nuclear spins were approximated by dipole-dipole interactions. The ZFS tensor was computed with the PyZFS code~\cite{ma_pyzfs_2020} using wavefunctions obtained from ground-state DFT calculations at the DDH level of theory and neglecting SOC effects. Although core electrons are not treated explicitly, for NVs in bulk diamond, the axial component of the ZFS tensor computed using the SG15 ONCV pseudopotentials has been shown to agree with PAW and experimental values within 5\%, likely due to error cancellations~\cite{seo_designing_2017}. The computed axial and transverse components of the ZFS tensor are presented in Fig.~\ref{s-fig:ZFS}.

\begin{table}[!ht]
\centering
\caption{
\textbf{Hyperfine interaction parameters (in MHz) for representative NV defects near the dislocation core and NV in bulk diamond (NV@Bulk).}
We present the results for the N atom and three C atoms in the closest proximity to the NV defect; the three C atoms are denoted as C1, C2, and C3 in the order of their increasing distance to the N atom. We considered nuclear isotopes $^{14}$N and $^{13}$C.
For each atom, we diagonalized the associated hyperfine $\bf{A}$-tensor, and obtained the eigenvalues. In the table, we list the three eigenvalues in ascending order for the $^{14}$N atom, and the largest eigenvalue for the $^{13}$C1, $^{13}$C2, and $^{13}$C3 atoms, respectively.
See Fig.~1 for the NV notations.}
\footnotesize
\begin{tabular}{lccccc}
\hline
Dislocation & NV defect & $^{14}$N & $^{13}$C1 & $^{13}$C2 & $^{13}$C3\\
\hline
None & NV@Bulk & (-2.14, -2.12, -1.72) & 178.40 & 176.94 & 176.64 \\
\hline
\multirow{9}{*}{30$^{\circ}$}
& G9-G8 & (-0.28, -0.27, 0.60) & 178.43 & 183.00 & 170.74 \\
& G3-G8 & (0.35, 0.52, 1.29) & 157.28 & 127.29 & 120.92 \\
& G12-G$^+$7 & (5.59, 5.69, 9.70) & 199.56 & 176.60 & 144.79 \\
& G10-G11 & (-0.12, -0.06, 1.71) & 166.87 & 143.24 & 166.62 \\
& G$^+$7-G$^+$3 & (-0.61, -0.56, 0.43) & 201.01 & 172.11 & 165.44 \\
& G9-G5 & (0.35, 0.44, 2.51) & 195.03 & 167.12 & 170.53 \\
& G13-G12 & (4.94, 5.07, 7.38) & 233.95 & 174.44 & 140.46 \\
& G11-G$^-$9 & (-0.68, -0.68, 0.62) & 204.49 & 178.67 & 162.48 \\
& G$^-$9-G$^-$10 & (-0.97, -0.94, -0.50) & 163.10 & 189.58 & 182.57 \\
\hline
\multirow{9}{*}{90$^{\circ}$}
& G4-G9 & (7.37, 7.49, 10.86) & 190.32 & 156.25 & 162.91 \\
& G10-G9 & (12.84, 12.94, 18.70) & 185.29 & 143.11 & 140.61 \\
& G16-G11 & (4.13, 4.29, 7.50) & 204.39 & 158.73 & 158.86 \\
& G$^+$9-G$^+$4 & (11.31, 11.38, 18.29) & 202.64 & 144.84 & 152.04 \\
& G5-G4 & (1.66, 1.78, 3.09) & 179.73 & 166.19 & 119.01 \\
& G15-G$^+$10 & (4.46, 4.51, 7.57) & 196.88 & 155.80 & 155.31 \\
& G17-G20 & (0.32, 0.40, 1.04) & 172.84 & 171.21 & 169.62 \\
& G$^+$1-G$^+$3 & (-0.75, -0.74, -0.19) & 171.64 & 161.12 & 167.33 \\
& G$^+$12-G$^+$13 & (-0.66, -0.62, 0.89) & 185.33 & 160.03 & 125.85 \\
\hline
\end{tabular}
\label{s-tab:hyperfine}
\end{table}

\section{Cluster correlation expansion calculations}

To study the spin coherence of NV center defects near dislocation cores, we carried out generalized cluster correlation expansion (gCCE) calculations with Monte Carlo bath state sampling using the PyCCE package~\cite{onizhuk_pycce_2021}. The coherence time is computed from the dynamics of the NV center initialized in $\ket{x} = \frac{1}{\sqrt{2}}\Bigl(\ket{0} \pm \ket{+}\Bigr)$ in a nuclear spin-bath environment. Under an applied magnetic field aligned with the NV quantization axis, the system is allowed to evolve freely for a duration $\tau$ (the inter-pulse delay and total free-evolution time used in $T^*_2$ measurements). The dephasing time $T^*_2$ is extracted from the free induction decay as a function of $\tau$. The Hahn-echo coherence time $T_2$ and the Carr-Purcell-Meiboom-Gill (CPMG) coherence time $T^{\rm CPMG}_2$ are obtained using pulse sequences of the form $(\tau - \pi - \tau)^N$, where $N$ is the number of $\pi$ pulses in the sequence. Simulations of $T_2^*$ and $T_2$ were carried out using second-order gCCE with 200 randomly sampled pure bath states (i.e., performing gCCE calculations for each pure bath state to approximate the mixed bath). For $T_2^{\rm CPMG}$ with $N \ge 8$, we employed sixth-order gCCE. The spin bath comprised naturally abundant $^{13}$C nuclei within a 5 nm cutoff radius, with a maximum inter-spin distance of 0.6 nm used in the cluster expansion.

\begin{figure}[!ht]
\centering
\includegraphics[width=0.7\textwidth]{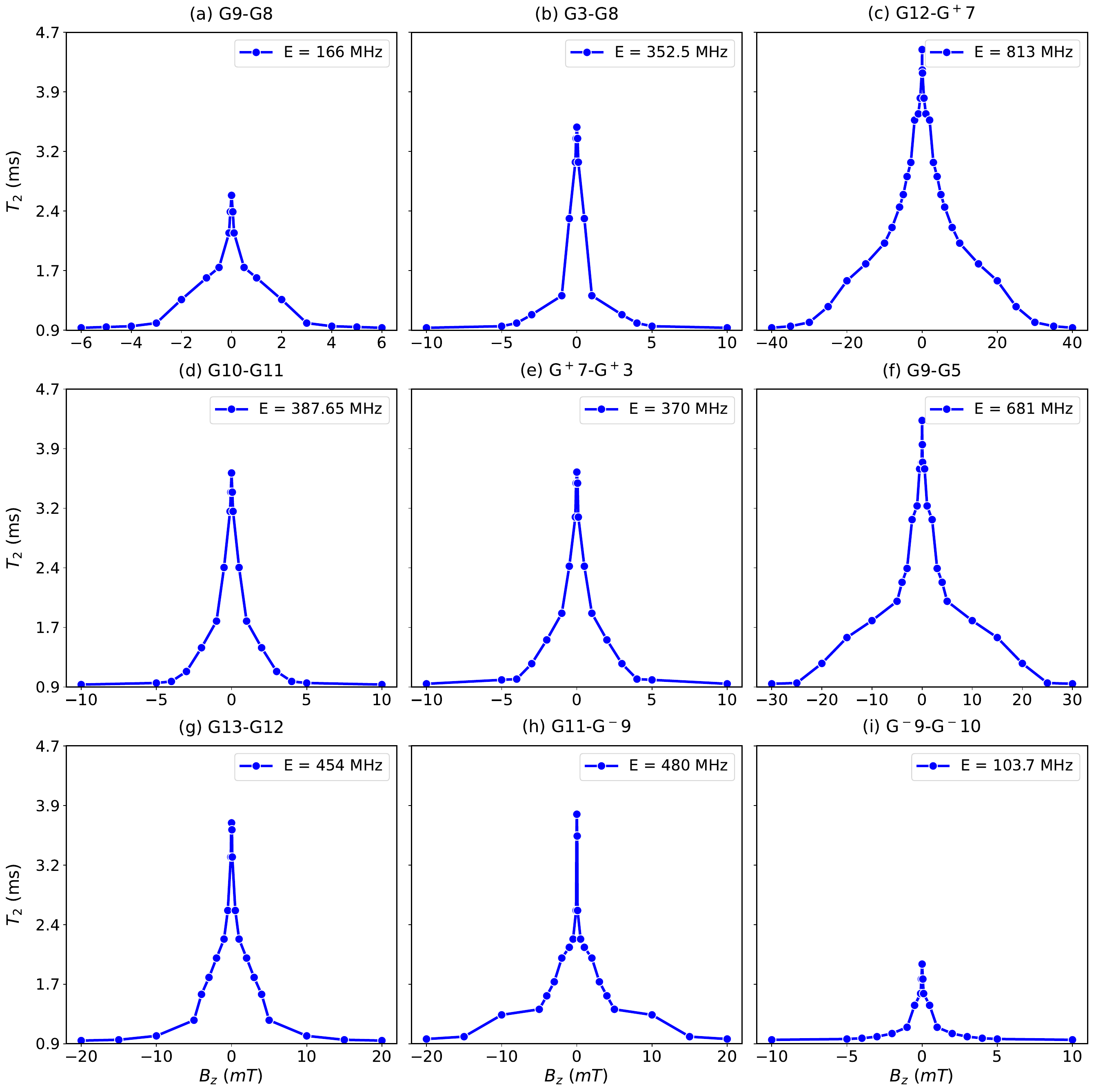}
\caption{
\textbf{Ensemble-averaged Hahn-echo coherence time ($T_2$) as a function of the magnetic field ($B_{z}$) for representative NV defects near the 30$^{\circ}$ dislocation core.}
The $T_2$ values peak at zero magnetic field, indicating that NV spin coherence is effectively preserved in the absence of an external field. A noticeable increase in $T_2$ is observed for NV centers with higher transverse ZFS ($E$) values, suggesting the improved coherence protection from magnetic noise for configurations with a larger $E$ at weak magnetic fields.
See Fig.~1 for the NV notations.}
\label{s-fig:30DP_CT}
\end{figure}

\begin{figure}[!ht]
\centering
\includegraphics[width=0.7\textwidth]{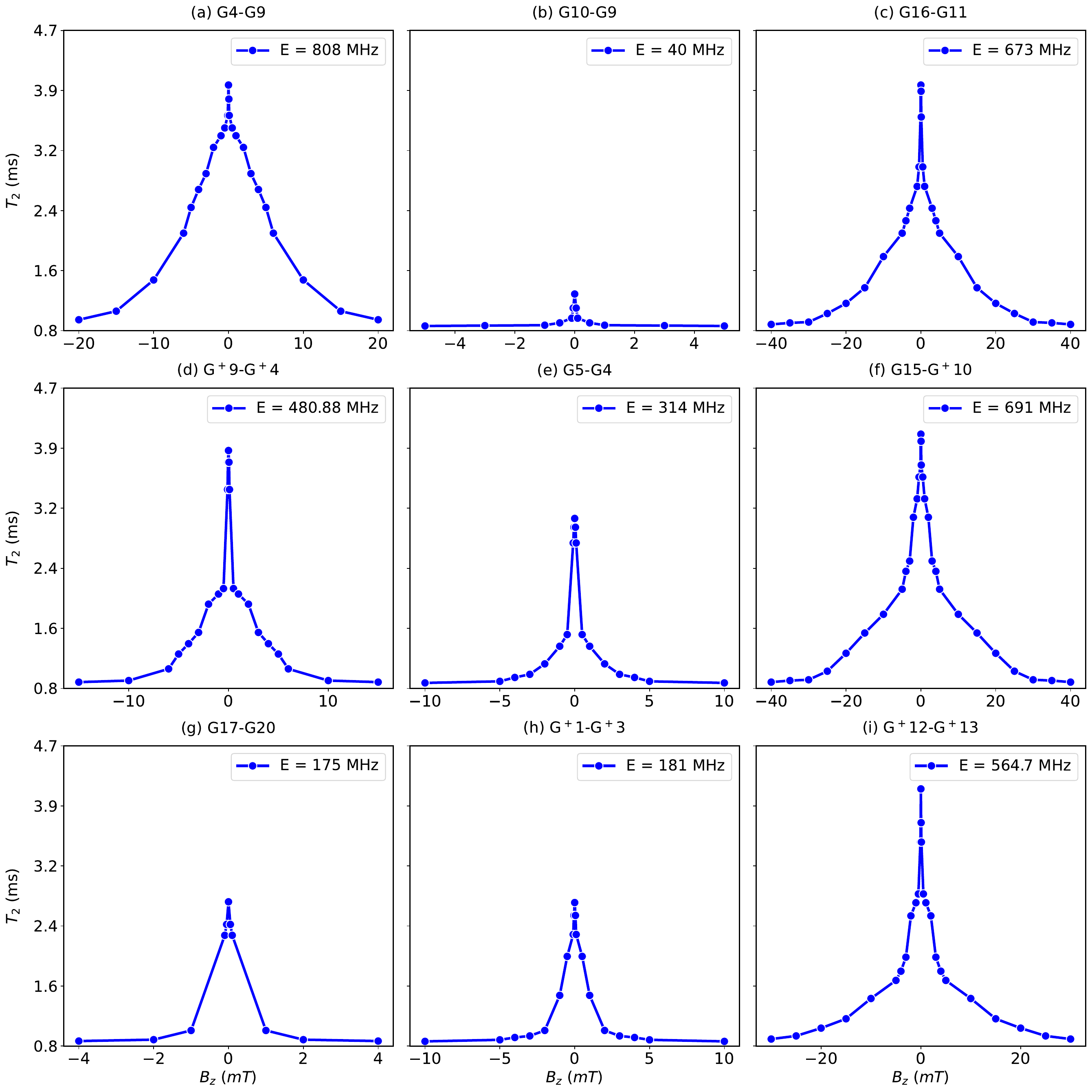}
\caption{
\textbf{Ensemble-averaged Hahn-echo coherence time ($T_2$) as a function of the magnetic field ($B_{z}$) for representative NV defects near the 90$^{\circ}$ dislocation core.}
The $T_2$ values peak at zero magnetic field, indicating that NV spin coherence is effectively preserved in the absence of an external field. A noticeable increase in $T_2$ is observed for NV centers with higher transverse ZFS ($E$) values, suggesting the improved coherence protection from magnetic noise for configurations with a larger $E$ at weak magnetic fields.
See Fig.~1 for the NV notations.}
\label{s-fig:90DP_CT}
\end{figure}

\begin{figure}[!ht]
\centering
\includegraphics[width=0.7\textwidth]{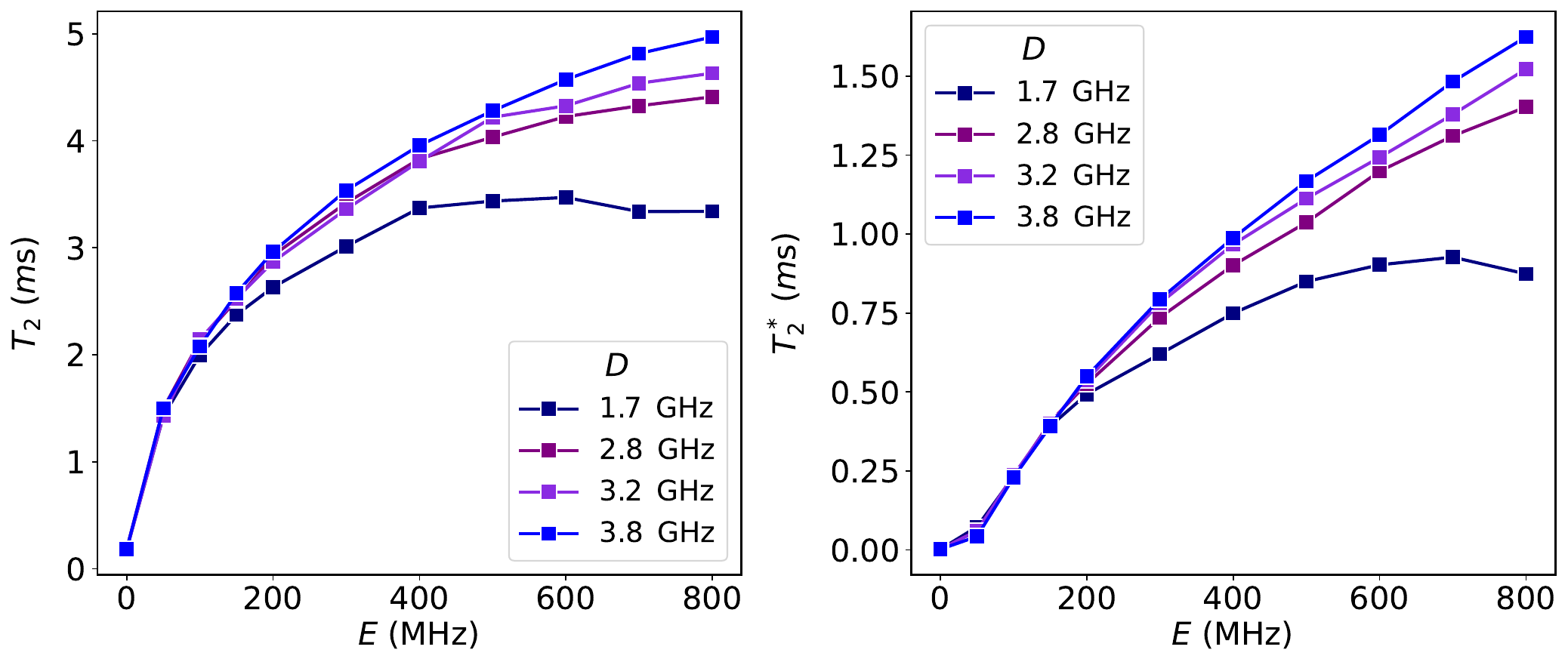}
\caption{
\textbf{Simulated $T_2$ and $T^*_2$ times at zero magnetic field as a function of transverse ZFS ($E$), considering different axial ZFS ($D$) parameters.}
Both $T_2$ and $T^*_2$ at zero magnetic field depend on the $E$ and $D$ values of the ZFS tensor.}
\label{s-fig:T2_ZFS}
\end{figure}

Given that nuclear spin flips at weak magnetic fields can arise from both hyperfine coupling and dipolar-dipolar interactions~\cite{zhao_decoherence_2012}, we used accurate hyperfine tensors from \textit{ab-initio} calculations for all $^{13}$C nuclei within our supercell, and for spins lying outside the supercell (beyond $\sim20$ \AA\ from the NV), we employed a point dipole approximation for their hyperfine coupling in our spin-dynamics simulations to compute the coherence times. All the results were obtained by sampling the NV center spin decoherence in 500 bath positional configurations (a bath configuration is a set of random positions of $^{13}$C atoms in the diamond lattice). For the representative NV configurations at 30$^{\circ}$ and 90$^{\circ}$ dislocations, we computed their Hahn-echo coherence times $T_2$ as a function of magnetic field ($B_z$). The results (Figs.~\ref{s-fig:30DP_CT} and~\ref{s-fig:90DP_CT}) show that all NV configurations near a dislocation exhibit increased $T_2$ at zero magnetic field due to non-zero transverse $E$ ZFS components (Fig.~\ref{s-fig:ZFS}), with larger $E$ yielding higher and more robust $T_2$ across the entire magnetic field range. To isolate the influence of the ZFS parameters, we took a single representative NV ($G^+7-G^+3$) with its full \textit{ab initio} hyperfine tensors and, keeping everything else fixed, systematically varied only $D$ and $E$. As shown in Fig.~\ref{s-fig:T2_ZFS}, we find that both the zero-field Hahn-echo coherence time $T_2$ and the free-induction decay time $T_2^*$ increase with $D$ and $E$, so that NV centers with larger longitudinal and transverse ZFS values exhibit substantially greater protection from magnetic noise.

\clearpage
\bibliographystyle{elsarticle-num}
\bibliography{disloc}